\documentclass{amsart}
\usepackage[english]{babel}
\usepackage{graphicx}
\usepackage[hidelinks]{hyperref} 
\usepackage{amssymb}
\usepackage{amsmath}
\usepackage[foot]{amsaddr}
\usepackage{amsfonts}
\usepackage{xcolor}
\usepackage{mathtools}
\usepackage{algpseudocode}
\usepackage{algorithm}
\usepackage[top=1in, bottom=1.25in, left=1.25in, right=1.25in]{geometry}

\graphicspath{{images/}}
\usepackage{float}
\usepackage{subcaption}

\newcommand{\bmu}{{\boldsymbol{\mu}}}
\newcommand{\V}{{\mathbb{V}}}
\newcommand{\X}{{\mathbb{X}}}
\newcommand{\Q}{{\mathbb{Q}}}
\newcommand{\N}{{\mathcal{N}}}
\newcommand{\R}{{\mathbb{R}}}
\DeclareMathOperator*{\argmax}{arg\,max}
\newcommand{\eqdot}{{\,\vcentcolon=\,}}
\newcommand{\Pa}{\mathcal{P}}
\newcommand{\oOm}{\overline{\Omega}}
\newcommand{\Om}{\Omega}
\newcommand{\norm}[1]{\left\lVert#1\right\rVert}
\newcommand{\abs}[1]{\left\lvert#1\right\rvert}

\begin{document}
\title[An ANN approach to bifurcating phenomena in CFD]{An artificial neural network approach to bifurcating phenomena in computational fluid dynamics}

\author{Federico Pichi$^{1,2}$}
\author{Francesco Ballarin$^3$}
\author{Gianluigi Rozza$^1$}
\author{Jan S. Hesthaven$^2$}
\address{$^1$ mathLab, Mathematics Area, SISSA, Trieste, Italy}
\address{$^2$ Chair of Computational Mathematics and Simulation Science, \'Ecole Polytechnique F\'ed\'erale de Lausanne, 1015 Lausanne, Switzerland}
\address{$^3$ Department of Mathematics and Physics, Catholic University of the Sacred Heart, Brescia, Italy}
\begin{abstract}
This work deals with the investigation of bifurcating fluid phenomena using a reduced order modelling setting aided by artificial neural networks. We discuss the POD-NN approach dealing with non-smooth solutions set of nonlinear parametrized PDEs. Thus, we study the Navier-Stokes equations describing: (i) the Coanda effect in a channel, and (ii) the lid driven triangular cavity flow, in a physical/geometrical multi-parametrized setting, considering the effects of the domain's configuration on the position of the bifurcation points. Finally, we propose a reduced manifold-based bifurcation diagram for a non-intrusive recovery of the critical points evolution. Exploiting  such detection tool, we are able to efficiently obtain information about the pattern flow behaviour, from symmetry breaking profiles to attaching/spreading vortices, even at high Reynolds numbers.  
\end{abstract}

\maketitle

\section{Introduction and motivation}

Nowadays, many authors are investigating the benefits that a machine learning (ML) \cite{Goodfellow-et-al-2016} approach could bring to different topics in numerical analysis \cite{2019JCoPh.378..686R,guo2019data,kutyniok2019theoretical,lee2020model,guo2018reduced,regazzoni2019machine,wang2019non}. On one hand, the higher computational resources and the increasing data available make the application of these methodologies easier and faster, but on the other hand it is clear that without relying on a solid mathematical analysis they could led to unreliable results.
In the context of the approximation of partial differential equations (PDEs), an example of such a situation is given by the existing duality between projection-based and data-driven methods \cite{benner2015,peherstorfer2015dynamic}. While the former builds upon the mathematical formulation of the problems (e.g.\ the weak formulation of the PDE), the latter only exploits the information from data to learn the dynamics.

The scientific community is making a great effort focusing the attention on a better mathematical characterization of the approximation properties of data-driven techniques based on the Neural Networks (NNs) \cite{cybenko89,shin2020convergence,Mishra2020EstimatesOT}. Despite this, results are still limited.

Nevertheless, projection-based methods are usually characterized by high computational burden and their applicability is compromised when a real-time or many-query context is needed. This is the case for the investigation of bifurcating phenomena \cite{seydel,Kuzne,caloz1997numerical} modelled by parametrized nonlinear PDEs, for example by means of high-fidelity discretization techniques such as the Finite Element (FE) method (or Finite Volume, Finite Difference, Spectral Element) \cite{quarteroni2008numerical}. The aim, when modelling such complex problems, is to detect and properly follow the bifurcating behaviour of the model, i.e.\ study the sudden qualitative changes and stability properties of the solutions w.r.t.\ a smooth variation of the bifurcating parameter. Indeed, the general setting includes a non-differentiable evolution of the solution manifold, and thus the non-uniqueness of the state.

This lead us to the investigation of the many possible existing solutions of a multi-parametrized system, where both physical and geometrical quantities are varied. Therefore, repeated evaluations of the PDE solutions, corresponding to the sampling over the parameter space, are required to obtain information about the model's behaviour.
In addition to this, fine mesh discretization have to be considered to be able to capture all effects of the nonlinearities in the system. Thus, to properly reconstruct the bifurcating behaviour, we need a collection of methodologies (e.g.\ continuation techniques \cite{allogwer}, deflation methods \cite{classic_deflation}, and eigenvalue problems \cite{pichi_bif}) which make such task computationally unbearable.

Recently, due to the high computational cost discussed above, a growing effort was put towards the use of Reduced Order Models (ROMs) \cite{benner2017model} to these bifurcating systems \cite{pichi_phd,pichistrazzullo,pichi_schro,pichi_bif,pintore2019efficient,pichipatera,HESS2019379, Herrero2013RBB,Terragni:2012}. The ultimate goal of ROMs is to build a low-dimensional manifold which: (i) is able to represent the physics involved, (ii) enables fast evaluation of the solution for new parameters, and (iii) provides an accurate approximation of the high-fidelity manifold.

Among the many possible ROM techniques, we focus on the Reduced Basis (RB) method \cite{hesthaven2015certified,QuarteroniManzoniNegri2015,patera07:book}, which takes advantage of the offline-online paradigm to build a reduced space from a collection of high-fidelity solutions (time-demanding phase to be computed on high performance resources), to obtain reduced coefficients, i.e.\ the coefficients appearing in the linear expansion w.r.t.\ the RB functions. In fact, the latter can be recovered by Proper Orthogonal Decomposition (POD) and span the reduced space.

Unfortunately, this offline-online decomposition is prevented in our setting by the nonlinear terms appearing in the PDEs, and from which the bifurcating phenomena arise. Therefore, the reduction of nonlinear parametrized PDEs usually still suffers from the curse of dimensionality and requires the adoption of intrusive hyper-reduction techniques such as the Empirical Interpolation Method (EIM) and its variants \cite{barrault04:_empir_inter_method,Chaturantabut2010}. These methodologies prevent the online phase to depend on the large number of degrees of freedom that characterizes the high-fidelity simulation, enabling a good speed-up.

In the bifurcating context, this intrusive hyper-reduction approach not only increases the computational burden during the offline phase, but it is also not robust due to the non-uniqueness of the solution \cite{pichi_phd,pichi_schro}.
To properly investigate the complex physics involved, we seek to adopt a method combining the reliability of the projection-based RB and the efficiency of the data-driven NNs.

Thus, we chose the POD-NN method presented and further investigated in \cite{hesthaven2018non,hestpinn} as the regression based non-intrusive RB method.
This approach relies on the combination of the POD technique, based on the Galerkin-FE method, and artificial neural networks. It enables a complete offline-online decoupling, recovering the efficiency without the need  of intrusive methodologies, while still relying on accurate high-fidelity approximations.

In practice, instead of retrieving the RB coefficients by means of the classic POD-Galerkin approach, where we have to intrusively assemble and solve at each step a nonlinear reduced system, we aim at training a feedforward neural network, based on the same POD dataset, to build a map from the parameter space to the reduced coefficients themselves. Other ML approaches could have been used to investigate this framework \cite{kalia2021learning,2019JCoPh.378..686R,fresca2020comprehensive,Brunton3932}, but we focus only on this POD-NN technique.

We analyze two relevant test cases in computational fluid-dynamics (CFD), where both physical and geometrical parameters have been taken into consideration. The first problem describes a viscous, steady and incompressible flow in a planar straight channel with a sudden expansion, while the second one deals with the driven flow inside a triangular cavity.

Indeed, both applications are subject to bifurcating phenomena: while the former models a cardiac disease, called mitral-valve regurgitation, caused by the \textit{Coanda effect} \cite{Quaini,pichistrazzullo,pintore2019efficient,HESS2019379}, the latter involves many different interesting behaviours that, to our knowledge, have not been studied in detail previously.

Moreover, since the computational domain has  great relevance for these applications, i.e.\ the steady state solutions and their uniqueness strongly depend on it, we encode parametrized geometries in the models to investigate how the uniqueness regime vary with it.

Thus, the goal is to exploit the regression based non-intrusive RB method to develop an efficient and deep understanding of complex bifurcating phenomena in the physically and geometrically parametrized context.

The standard way to study the solutions' behaviour is the\textit{ bifurcation diagram}, which is a tool that exploits specific features of the model to understand the evolution of the solutions manifold w.r.t.\ the bifurcating parameter. It is possible to build an affine output directly from the reduced solution to obtain a representation of the bifurcation, but this would still depend on the intrusive RB phase. Motivated by this, we develop a reduced manifold based bifurcation diagram (RMBBD). The latter is a detection tool which exploits the parameter dependence of the RB coefficients to discover, in a non-intrusive manner, the evolution of the critical points corresponding to the branching behaviour.

All simulations were performed within the open source software FEniCS \cite{fenics} and RBniCS \cite{rbnics}, while we chose PyTorch \cite{pytorch} to construct the neural network.

The work is organized as follows: in Section \ref{sec:approximation} we present the general nonlinear parametrized setting for a PDE undergoing bifurcating phenomena, in both the continuous and the discrete context.
In Section \ref{sec:reduced} we briefly introduce the reduced order methodology that will serve to cut the computational burden and to provide the low fidelity space needed for the non-intrusive investigation. Then, in Section \ref{sec:ann} we discuss the POD-NN approach for an efficient recovery of the reduced coefficients and introduce the reduced manifold based bifurcation diagram that we develop for a non-intrusive detection of the bifurcation points. In Section \ref{sec:results} we discuss the behaviour of the Navier-Stokes model and the RB/POD-NN results for: (i) the Coanda effect in a channel, and (ii) the lid driven triangular cavity flow, both of them in a multi-parametrized context. Finally, in Section \ref{sec:conc} we draw some conclusions and highlight future perspectives.

\section{Approximation of bifurcating PDEs}\label{sec:approximation}

Let us consider a generic parametrized PDE which models a system whose behaviour is affected by some physical or geometrical parameters. When the parameters change slightly, we would expect the solution to evolve continuously in a unique manner. This may not be the case when dealing with complex phenomena modelled by nonlinear equations. Indeed, sudden changes in the qualitative solutions' behaviour can occur at critical parameter values, known as the \textit{bifurcation points} \cite{seydel,ambrosetti1995primer}. 
Moreover, when reaching these values, we may loose the uniqueness of the solution for a given configuration of the system, i.e.\ a specific value of the parameter. 
We will refer to the set of solutions with the same qualitative properties as a \textit{branch}.
Thus, an extensive numerical study of such models is required to understand the underlying physics. 

Denoting the parameter as $\boldsymbol{\mu} \in \mathcal{P} \subset \mathbb{R}^P$, the strong form of a parametrized PDE, over the original (parameter-dependent) domain $\Omega(\bmu) \subset \mathbb{R}^d$, can be expressed as 
\begin{equation}
\label{eq:strong_form}
G(X(\boldsymbol{\mu}); \boldsymbol{\mu}) = 0 \qquad \text{in } \mathbb{X}'_{\bmu} \, ,
\end{equation}
where $X(\bmu) \in \mathbb{X}_{\bmu}$ is the solution belonging to some suitable Hilbert space $\X_{\bmu} \eqdot \X(\Omega(\bmu))$, and $G: \X_{\bmu} \times \mathcal{P} \to \X'_{\bmu}$ is the parametrized map representing the PDE.

The numerical approximation of these problems often involves their weak formulation, that reads as: given $\boldsymbol{\mu} \in \mathcal{P}$, find $X(\boldsymbol{\mu}) \in \X_{\bmu}$ such that
\begin{equation}
\label{eq:weak_form}
g(X(\boldsymbol{\mu}), Y; \boldsymbol{\mu}) \eqdot \langle G(X(\boldsymbol{\mu}); \boldsymbol{\mu}), Y\rangle_{\X'_{\bmu} , \X_{\bmu}} = 0 \qquad \forall \ Y \in \X_{\bmu} \, ,
\end{equation}
where we have introduced the parametrized variational form $g: \X_{\bmu} \times \X_{\bmu} \times \mathcal{P} \to \mathbb{R}$.

Let $(Z; \bmu) \in \X_{\bmu} \times \Pa$ be a known solution of \eqref{eq:strong_form}. Then, to ensure the well-posedness of the problem, one usually assumes that the Frech\'et partial derivative $D_XG(Z; \bmu): \mathbb{X}_{\bmu} \rightarrow  \mathbb{X}'_{\bmu}$ is bijective (thanks to the Implicit Function Theorem \cite{ciarlet2013linear,caloz1997numerical}).

Introducing the parametrized variational form $\mathrm{d}g[Z](\cdot, \cdot; \bmu)$ expressing the Frech\'et partial derivatives of $G$ w.r.t.\ $X$ at $(Z, \bmu) \in \X_{\bmu} \times \Pa$ as
\begin{equation}
\label{eq:var_form_der}
\mathrm{d}g[Z](X, Y; \bmu) \eqdot \langle D_XG(Z; \bmu)X, Y\rangle \quad \forall \, X, Y \in \X_{\bmu}  \, ,
\end{equation}
the bijectivity can be reformulated in terms of the variational form \eqref{eq:var_form_der}, as:
\begin{itemize}
\item[$\circ$] \textit{continuous} on $\X_{\bmu} \times \X_{\bmu}$, if there exists a continuity constant $\overline{\gamma} > 0$ such that
\begin{equation}
\label{eq:continuity}
\gamma(\bmu) \eqdot  \sup_{X \in \X_{\bmu}} \sup_{Y \in \X_{\bmu}} \frac{\mathrm{d}g[Z](X, Y; \bmu)}{\norm{X}_{\X_{\bmu}}\norm{Y}_{\X_{\bmu}}} > \overline{\gamma} \qquad \forall \, \bmu \in \Pa \, ,
\end{equation}
\item[$\circ$] \textit{inf-sup stable} on $\X_{\bmu} \times \X_{\bmu}$, if there exists an inf-sup constant $\overline{\beta} > 0$ such that
\begin{equation}
\label{eq:inf-sup_1}
\beta(\bmu) \eqdot  \inf_{X \in \X_{\bmu}} \sup_{Y \in \X_{\bmu}} \frac{\mathrm{d}g[Z](X, Y; \bmu)}{\norm{X}_{\X_{\bmu}}\norm{Y}_{\X_{\bmu}}} \geq \overline{\beta} \qquad \forall \, \bmu \in \Pa \, ,
\end{equation}
and
\begin{equation}
\label{eq:inf-sup_2}
 \inf_{Y \in \X_{\bmu}} \sup_{X \in \X_{\bmu}} \frac{\mathrm{d}g[Z](X, Y; \bmu)}{\norm{X}_{\X_{\bmu}}\norm{Y}_{\X_{\bmu}}} > 0 \qquad \forall \, \bmu \in \Pa \, .
\end{equation}
\end{itemize}

Since the conditions \eqref{eq:inf-sup_1} and \eqref{eq:inf-sup_2} are equivalent, respectively, to the injectivity and the surjectivity of the Frech\'et derivative $D_XG(Z; \bmu)$, the existence of a \textit{local branch of non-singular solutions} is guaranteed.

When the parameter reaches a critical value $\bmu^*$, i.e.\ the bifurcation point, the inf-sup stability of the model is lost, and the system admits the existence of a qualitatively different solution that bifurcates from the previous branch. Indeed, the inf-sup constant $\beta(\bmu^*)$  becomes zero, and the Frech\'et derivative $D_XG(Z; \bmu)$ fails to be invertible.

The main goal of \textit{bifurcation theory} is to provide a mathematical description of the bifurcating scenario that can be observed in physical systems.
The understanding of the global behaviour is usually pursued by means of the reconstruction of the bifurcation diagram, i.e.\ the plot of a scalar characteristic output $s(X(\bmu))$, corresponding to a given solution $(X(\bmu), \bmu)$ of the PDE, for each instance of the parameter in $\Pa$.
Therefore, the existence of qualitatively different solutions for the same values of the parameter will result in the presence of multiple branches. 

The bifurcating phenomena depend on the physical domain in which the problem is posed. Therefore, to pursue a more general investigation, we considered a parametrized geometry $\Om(\bmu)$. 
A more practical way to deal with \eqref{eq:weak_form} is to express the weak formulation over a reference (parameter-independent) domain $\oOm \subset \mathbb{R}^d$, which is usually taken as $\oOm = \Om(\overline{\bmu})$ with $\overline{\bmu} \in \mathcal{P}$.
Finally, denoting by $\boldsymbol{\Phi}: \oOm \times \Pa \to \Om(\bmu)$ the (affine) transformation map such that $\Omega(\bmu) = \boldsymbol{\Phi}(\oOm, \bmu)$, we can pull back the weak formulation \eqref{eq:weak_form} over the reference domain $\oOm$ as follows: given $\boldsymbol{\mu} \in \mathcal{P}$, find $X(\boldsymbol{\mu}) \in \X$ such that
\begin{equation}
\label{eq:weak_form_ref}
g(X(\boldsymbol{\mu}), Y; \boldsymbol{\mu}) = 0 \qquad \forall \ Y \in \X \, ,
\end{equation}
where, with a little abuse of terminology, we have kept the same notation for all quantities now defined on the reference domain $\oOm$.
Furthermore, the solution over the original domain $\Om(\bmu)$ can be recovered by means of the composition $X(\bmu) \circ \boldsymbol{\Phi}(\cdot \, ; \bmu)$. 

For the numerical approximation of \eqref{eq:weak_form_ref}, a full-order (or high-fidelity) discretization method is needed. Here we focus on the Finite Element method \cite{quarteroni2008numerical}, which exploits a Galerkin projection of the system over a finite dimensional subspace $\X_{\N} \subset \X$ of dimension $\N$.

The Galerkin-FE method reads as: given $\bmu \in \Pa$, we seek $X_\mathcal{N}(\bmu) \in \X_\mathcal{N}$ that satisfies 
\begin{equation}
\label{eq:weakgal}
g(X_\mathcal{N}(\bmu), Y_\mathcal{N} ; \bmu) = 0\ , \quad \forall \, Y_\mathcal{N} \in \X_\mathcal{N} \, . 
\end{equation}
The nonlinear solver chosen to linearize the weak formulation in \eqref{eq:weakgal} is the Newton-Kantorovich method \cite{ciarlet2013linear,QuarteroniManzoniNegri2015}, given as: for $\bmu \in \Pa$ and an initial guess $X_{\N}^0(\bmu) \in \X_{\N}$, for every $k = 0, 1, \dots$, we seek the variation $\delta X_{\N} \in \X_{\N}$ such that
\begin{equation}
\label{eq:weaknewton}
\mathrm{d}g[X_{\N}^k(\bmu)](\delta X_{\N}, Y_{\N}; \bmu) =  g(X_{\N}^k(\bmu),Y_{\N} ; \bmu) \ , \quad \forall \, Y_{\N} \in \X_{\N} \, .
\end{equation}
We update the solution as $X_{\N}^{k+1}(\bmu) = X_{\N}^{k}(\bmu) - \delta X_{\N}$, and repeat these steps until an appropriate stopping criterion is verified.


From the algebraic point of view, we denote with $\{E^j\}_{j=1}^{\mathcal{N}}$ a Lagrangian basis for $\X_{\mathcal{N}}$, such that, denoting the solution vector as ${\bf{X}}_{\N}(\bmu) = \{X_{\N}^{(j)}(\bmu)\}_{j=1}^{\N}$, we can write each element $X_\mathcal{N}(\bmu) \in \X_{\mathcal{N}}$ as 
\begin{equation}
\label{eq:soldecomp}
X_{\N}(\bmu) = \sum_{j=1}^{\N} X_{\N}^{(j)}(\bmu)E^j \, .
\end{equation}
Thus, we return to the study of the solution $\bf{X}_{\N}(\bmu) \in \R^{\N}$ of the system provided by the combination of 
the Newton-Kantorovich method's $k$-th step and the Galerkin-FE discretization: 
given $\bmu \in \Pa$ and an initial guess ${\bf{X}}_{\N}^0(\bmu) \in \R^{\N}$, for every $k = 0, 1, \dots$, we seek the variation $\delta {\bf{X}}_{\N} \in \R_{\N}$ such that
\begin{equation}
\label{eq:linearnewtgal}
\mathsf{J}_\N({\bf{X}}_{\N}^k(\bmu); \bmu) \delta{\bf{X}}_{\N} = \mathsf{G}_{\N}({\bf{X}}_{\N}^k(\bmu); \bmu) \ ,
\end{equation}
where the high-fidelity residual vector in $\R^\N$ and Jacobian matrix in $\R^{\N \times \N}$ are defined as 
\begin{equation}
\begin{aligned}
\label{eq:jacobiandef}
&(\mathsf{G}_{\N}({\bf{X}}^k_{\N}(\bmu); \bmu))_i = g({\bf{X}}^k_{\N}(\bmu), E^i ; \bmu),   \quad\quad \forall \,  i = 1, \dots, \mathcal N \, , \\ 
&(\mathsf{J}_\N({\bf{X}}_{\N}^k(\bmu); \bmu))_{ij} = \mathrm{d}g[{\bf{X}}_{\N}^k(\bmu)](E^j, E^i; \bmu) , \quad\quad \forall \,  i, j = 1, \dots, \mathcal N \, .
\end{aligned}
\end{equation}
Let us note that, even when considering a non-bifurcating context, one has to take care of the discrete counterpart of the inf-sup stability conditions. Indeed, a common choice for the FE space in computational fluid-dynamics applications, where velocity and pressure fields are involved, is the Taylor-Hood $\mathbb{P}^2 - \mathbb{P}^1$ FE discretization.

Thus, given $\bmu \in \Pa$ we are able to find an approximate solution $X_{\N}(\bmu)$ of \eqref{eq:weakgal}. However, we aim at investigating solution's behaviour following its branching phenomenon. For this reason, the presented framework needs to be coupled  with a \textit{continuation method} \cite{allogwer,seydel,Kuzne}, which enables the convergence to a prescribed branch.

Finally, we remark that the computational cost of a deep analysis for a given bifurcating model is often infeasible. In fact, we should solve the possibly large linear system \eqref{eq:linearnewtgal} in a nested way; i.e.\ , for every iteration of the Newton-Kantorovich method, for every parameter in the continuation method and for every solution branch that we want to reconstruct.
To enable this, we introduce in the next section a reduced order modelling approach to recover the bifurcation diagram in a more efficient way.

\section{Reduced order model}\label{sec:reduced}

Dealing with nonlinear parametrized PDEs remains a computational challenge. During the last decade, much efforts has been put towards Reduced Order Models \cite{benner2017model}, a class of techniques that are very well suited when dealing with many-query or real time investigations.

One of the main techniques is the Reduced Basis method \cite{hesthaven2015certified,QuarteroniManzoniNegri2015,patera07:book}, which enables a fast evaluation of the solution for a new parameter $\boldsymbol{\mu} \in \mathcal{P}$. This is possible thanks to the decomposition of such methodology into two phases, the offline and the online one. During the former, one precomputes the high-dimensional terms coming from the full-order approximation technique and then assembles all relevant quantities. On the contrary, during the latter, one rapidly solves a reduced system of much smaller dimension, where the efficiency comes from its independence from the degrees of freedom $\N$ of the underlying high-fidelity method.

Therefore, while the scope of the offline phase is the construction of a low dimensional basis
for the \textit{reduced manifold} $\mathbb X_N \subset \mathbb X_\mathcal{N}$, the online phase is the model resulting from the Galerkin projection onto this precomputed space, to obtain the solution as a linear combination of the RB functions which span $\mathbb X_N$.

Among many possible ways to construct a basis for the reduced manifold, we focus here on the Proper Orthogonal Decomposition \cite{volkwein2011model}, a singular value decomposition based algorithm exploited to extract the most meaningful information about the system. Even though it requires a greater computational effort during the offline phase (when compared with other sampling/construction strategies), this choice is motivated by: (i) better approximation capability  when dealing with the bifurcating context, (ii) no need for the development of an a posteriori error estimator (required instead by the Greedy algorithm \cite{refId0,BRR1}), (iii) good fit with the machine learning paradigm we will present later on, and (iv) easy combination with continuation strategies.

Given $\mathcal{P}_{\text{train}} = \{\boldsymbol{\mu}^{(i)}\}_{i=1}^{N_{\text{train}}} \subset \mathcal{P}$, a collection of parameters, this method entails solving the corresponding $N_{\text{train}}$ Galerkin-FE problems associated to each value in $\Pa_{\text{train}}$. Then, one builds the so-called \textit{snapshots} matrix $\mathsf{S}$ from the high-fidelity solutions vectors $\{{\bf{X}}_\N(\bmu^{(i)})\}_{i=1}^{N_{\text{train}}}$, which contain the information about the $\bmu$-dependence of the solutions manifold.

Finally, applying the POD algorithm to $\mathsf{S}$, one obtains the best basis of rank $N$ (among all the orthonormal ones) by taking the first $N$ left singular vectors. Denoting these basis functions as $\{\Sigma^n\}_{n=1}^{N}$, we can define the reduced space as $\X_N = \textrm{span}\{\Sigma^1, \dots, \Sigma^N\}$.

The starting point to obtain an efficient online phase is to express every element $X_N(\boldsymbol{\mu}) \in \X_{N}$ as a linear combination of the RB functions $\{\Sigma^n\}_{n=1}^{N}$, that is
\begin{equation}
\label{eq:rb_expansion}
X_N(\boldsymbol{\mu}) = \sum_{n=1}^N X_{N}^{(n)}(\boldsymbol{\mu}) \Sigma^n \, ,
\end{equation}
where the reduced solution ${\mathsf{X}}_{N}(\boldsymbol{\mu}) = \{X_{N}^{(n)}(\boldsymbol{\mu})\}_{n=1}^{N} \in \mathbb{R}^N$ is the coefficients vector in the RB expansion.

Here, we denote by $\boldsymbol{\Sigma}^n \in \mathbb{R}^{\N}$ the nodal values of the basis function $\Sigma^n$ for all $n =1, \dots, N$, and express with the matrix $\mathsf{V} = [\boldsymbol{\Sigma}^1 | \dots | \boldsymbol{\Sigma}^N] \in \mathbb{R}^{\N \times N}$ the change-of-variable  mapping from the RB basis to the FE basis, i.e.\ we write ${\bf{X}}_N = \mathsf{V}\mathsf{X}_{N}$.

During the online phase we project the linearized weak formulation \eqref{eq:weaknewton} onto the reduced space $\X_N$, obtaining the algebraic problem: given $\bmu \in \Pa$, and an initial guess ${\mathsf{X}}_{N}^0(\bmu) \in \R^{N}$, for every $k = 0, 1, \dots$, we seek the variation $\delta\mathsf{X}_{N} \in \R^{N}$ such that
\begin{equation}
\label{eq:weakred}
\mathsf{J}_N({\mathsf{X}}_{N}^k(\bmu); \bmu) \delta\mathsf{X}_{N} = \mathsf{G}_{N}({\mathsf{X}}_{N}^k(\bmu); \bmu) \, ,
\end{equation}
where $\mathsf{G}_N \in \R^{N}$ and $\mathsf{J}_N \in \R^{N \times N}$ are, respectively, the reduced residual and the reduced Jacobian matrix. Equation \eqref{eq:weakred} can be equivalently written as
\begin{equation}
\label{eq:jacobianreddef}
\mathsf{V}^T \mathsf{J}_\N(\mathsf{V}\mathsf{X}_N^k(\bmu); \bmu)\mathsf{V} \delta\mathsf{X}_{N} = \mathsf{V}^T\mathsf{G}_\N(\mathsf{V}\mathsf{X}_N^k(\bmu); \bmu) \, .
\end{equation}
We update the solution as $\mathsf{X}_N^{k+1}(\bmu) = \mathsf{X}_N^{k}(\bmu) - \delta \mathsf{X}_N$ until convergence.

Once again, let us remark that the inf-sup stability properties are not inherited from the underlying FE discretization step, and thus we need to recover such feature also at the reduced level. This task is accomplished by means of standard \textit{supremizer enrichment} \cite{bal}.

Moreover, it is clear that \eqref{eq:jacobianreddef} still involves the degrees of freedom $\N$ of the high-fidelity problem, thus a repeated assembly of the system compromises the efficiency of the RB during the online phase.
This issue can be overcome by adopting hyper-reduction techniques, such as the Empirical Interpolation Method (EIM) and its variants \cite{barrault04:_empir_inter_method,Chaturantabut2010}, to allow a consistent speed-up by recovering the affine dependency w.r.t.\ the parameter $\bmu$.
The application of such intrusive affine-recovery techniques (in the bifurcating context) is not robust (due to the presence of multiple existing solutions corresponding to a fixed $\bmu$) and adds further complexity to the offline phase \cite{pichi_phd,pichi_schro}.
The lack of efficiency of the methodology described above has brought us towards a non-intrusive investigation, that we explore in the next section.

\section{A non-intrusive POD-NN approach for bifurcating phenomena}
\label{sec:ann}
In this section, we briefly describe the POD-NN approach presented and further investigated in \cite{hesthaven2018non,hestpinn}, discussing how neural networks can be used to recover a non-intrusive reduced approximation of a generic nonlinear and parameter dependent PDE. In particular, we aim at building a regression model able to efficiently retrieve the RB coefficients to investigate branching phenomena, and eventually use them to obtain information about the location of the bifurcation points. 

The first step is the construction of the dataset used for the regression procedure. Having denoted with $\{E^{i}\}_{i=1}^{\N}$ a basis for the FE space $\X_\N$ and with $\{\Sigma^i\}_{i=1}^{N}$ a basis for the RB space $\X_N$, we recall that the $||\cdot||_{\X_\N}$ - closest element of $\X_N$ to the high-fidelity solution $X_\N$ can be expressed by its projection onto the reduced space itself as $$X^{\mathsf V}_\N(\boldsymbol{\mu}) \eqdot \sum_{j=1}^{\N} (\mathsf {V}\mathsf{V}^T{\bf{X}}_\N(\boldsymbol{\mu}))_jE^j = \sum_{i=1}^{N} (\mathsf{V}^T{\bf{X}}_\N(\boldsymbol{\mu}))_i\Sigma^i  \in \X_\N.$$
This means that, once we built the RB space by means of the POD technique, we can approximate the high-fidelity solution via a regression task that finds the ``best" RB coefficients in that basis.

We seek to build a neural network to approximate the function 
$\pi: \mathcal{P} \subset \mathbb{R}^P \to \mathbb{R}^N$, which maps each input parameter $\boldsymbol{\mu} \in \mathcal{P}$ to the vector of reduced coefficients  \begin{equation}
\label{eq:map_pi}
\pi(\boldsymbol{\mu}) \eqdot \mathsf{V}^T{\bf{X}}_\N(\boldsymbol{\mu}) \in \R^N
\end{equation}
expressing the expansion of $X_\N^\mathsf{V}(\boldsymbol{\mu})$ in the reduced basis $\{\Sigma^i\}_{i=1}^{N}$.

The learning procedure is performed through a supervised learning approach and is based on the training set given by the pairs $\{(\boldsymbol{\mu}^{(i)}, \pi(\boldsymbol{\mu}^{(i)})\}_{1 \leq i \leq N_{train}}.$
For the sake of consistency, rather then computing the training output for $\boldsymbol{\mu}^{(i)}$ as in \eqref{eq:map_pi}, we project the snapshots through the normal equations, solving the system given by $\mathsf{V}^T \mathsf{M} \mathsf{V} \pi(\bmu) = \mathsf{V}^T \mathsf{M} \bf{X}_\N(\bmu)$, where $\mathsf{M}$ is the inner product matrix of the discretization.
The main advantage of this training process is that it does not affect the offline phase of the reduction strategies. Indeed, one only has to rely on the computed snapshots to perform the POD compression step, i.e.\ : for any given $\bmu \in \Pa$, the POD-NN solution can be recovered non-intrusively by evaluating the regression map $\pi(\bmu)$ and projecting the vector of reduced coefficients in the high-fidelity space $\X_{\N}$, to obtain ${\bf{X}}_{\text{NN}}(\boldsymbol{\mu}) = \mathsf{V} \pi(\bmu) \in \R^\N$. Thus, denoting with $\mathsf{X}_{NN}(\bmu) \eqdot \pi(\bmu)$ the NN coefficients, the RB expansion of the POD-NN solution reads
\begin{equation}
\label{eq:nn_expansion}
{X}_{\text{NN}}(\boldsymbol{\mu}) = \sum_{n=1}^N X_{NN}^{(n)}(\boldsymbol{\mu}) \Sigma^n \,  \in \X_{\N} .
\end{equation}
We note that another advantage of the POD-NN non-intrusive approximation consists of the fact that we do not need to implement any continuation strategies. Indeed, just by querying the trained network, we obtain the reduced coefficients for any given value of $\bmu \in \Pa$ without the need for a proper initial guess to guide the convergence of the nonlinear solver. 

\begin{figure} [htb]   
\centering 
\def\svgwidth{8cm}   
\begingroup%
  \makeatletter%
  \providecommand\color[2][]{%
    \errmessage{(Inkscape) Color is used for the text in Inkscape, but the package 'color.sty' is not loaded}%
    \renewcommand\color[2][]{}%
  }%
  \providecommand\transparent[1]{%
    \errmessage{(Inkscape) Transparency is used (non-zero) for the text in Inkscape, but the package 'transparent.sty' is not loaded}%
    \renewcommand\transparent[1]{}%
  }%
  \providecommand\rotatebox[2]{#2}%
  \newcommand*\fsize{\dimexpr\f@size pt\relax}%
  \newcommand*\lineheight[1]{\fontsize{\fsize}{#1\fsize}\selectfont}%
  \ifx\svgwidth\undefined%
    \setlength{\unitlength}{497.8566948bp}%
    \ifx\svgscale\undefined%
      \relax%
    \else%
      \setlength{\unitlength}{\unitlength * \real{\svgscale}}%
    \fi%
  \else%
    \setlength{\unitlength}{\svgwidth}%
  \fi%
  \global\let\svgwidth\undefined%
  \global\let\svgscale\undefined%
  \makeatother%
  \begin{picture}(1,0.56525948)%
    \lineheight{1}%
    \setlength\tabcolsep{0pt}%
    \put(0,0){\includegraphics[width=\unitlength,page=1]{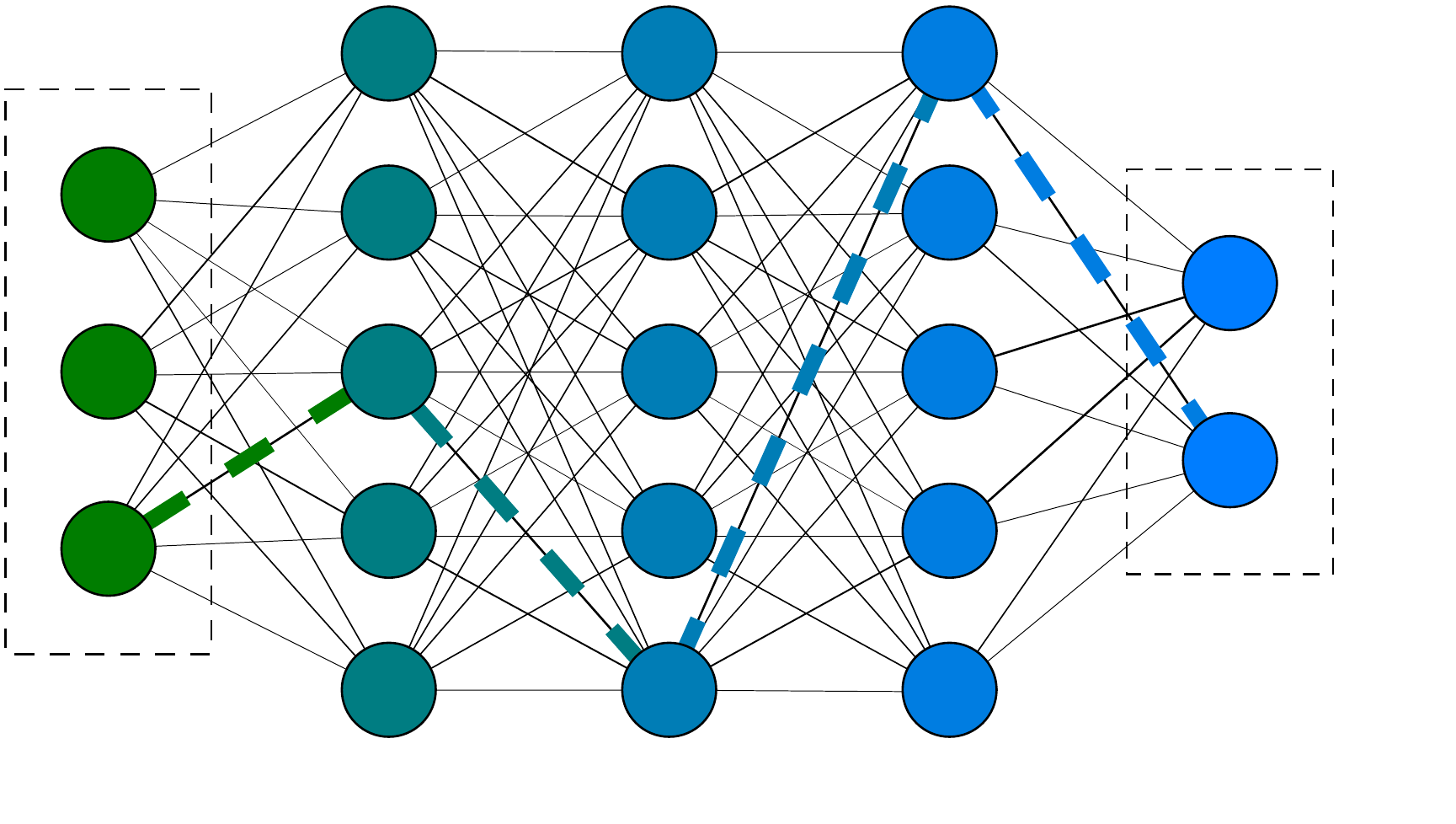}}%
    \put(0.04456693,0.03713517){\makebox(0,0)[lt]{\lineheight{1.25}\smash{\begin{tabular}[t]{l}$\mathbb{R}^P$\end{tabular}}}}%
    \put(0.8198826,0.08534181){\makebox(0,0)[lt]{\lineheight{1.25}\smash{\begin{tabular}[t]{l}$\mathbb{R}^N$\end{tabular}}}}%
    \put(0,0.54999879){\makebox(0,0)[lt]{\lineheight{1.25}\smash{\begin{tabular}[t]{l}\footnotesize{INPUT}\end{tabular}}}}%
    \put(0.73885149,0.49576626){\makebox(0,0)[lt]{\lineheight{1.25}\smash{\begin{tabular}[t]{l}\footnotesize{OUTPUT}\end{tabular}}}}%
    \put(0.26097229,0.00327575){\makebox(0,0)[lt]{\lineheight{1.25}\smash{\begin{tabular}[t]{l}\footnotesize{HIDDEN LAYERS}\end{tabular}}}}%
    \put(0,0){\includegraphics[width=\unitlength,page=2]{ann_net.pdf}}%
  \end{picture}%
\endgroup%

\caption{A feedforward neural network receiving the parameter $\bmu \in \Pa$ as input and returning the reduced coefficients $\pi(\bmu)$ as output, with $L_K=3$ layers and $H_K=5$ neurons.}
\label{fig:neuralnetwork}
\end{figure}

Despite the huge variety of neural network structures, we chose here a standard feedforward neural network, or multi-layer perceptron (MLP), shown in Figure \ref{fig:neuralnetwork}. The latter is an oriented graph, where the nodes, or \textit{neurons}, are organized as follows: (i) an input layer of dimension $P$ corresponding to the dimension of the parameter space $\Pa$; (ii) an output layer of dimension $N$ corresponding to the dimension of the RB space; (iii) $L_K$ inner layers, each with $H_K$ computing neurons. 
We set a learning rate $\eta$, the weighted sum as a propagation function, the hyperbolic tangent $\tanh(x)$ as activation functions, and the identity as output function in the last layer.

Moreover, once the dataset given by the parameters and the corresponding target outputs is obtained, we split it for cross-validation purposes as follows: a training set $\Xi_{tr} = \{\boldsymbol{\mu}^{(1)}, \dots, \boldsymbol{\mu}^{(N_{tr})}\}$ with $N_{tr} = (5/6) \cdot N_{train}$ and a validation set $\Xi_{va} = \{\boldsymbol{\mu}^{(1)}, \dots, \boldsymbol{\mu}^{(N_{va})}\}$ with $N_{va} = (1/6) \cdot N_{train}$.
Thus, during the training procedure, we are using ${\sim}15\%$ of the data to obtain an indicator, to gauge the generalizing capabilities of the network on the validation set.
We want to prevent \textit{overfitting}, which appears when the NN memorizes the training set exactly and thus produce bad approximation of unseen data.



As concerns the training procedure, we took into account both mini-batches and learning through the epochs paradigm. Within this setting, we adjust the weights of the network by means of the ADAM optimizer \cite{Kingma2015AdamAM}, guided by the mean squared error (MSE) function as an indicator of the performance. The MSE provides a measure of the discrete error obtained by replacing the projected solution $X^{\mathsf V}_\N(\boldsymbol{\mu})$ with the output of the neural network regression $X_{\text{NN}}(\boldsymbol{\mu})$. 
The ADAM method is a first order technique that considers only the gradient of the cost function, while for example the optimizer used in \cite{hesthaven2018non} is the Levenberg-Marquardt algorithm, which is of the second order and thus requires the computation of the Hessian. This makes our approach faster but less accurate. 


We initialized the weights with a uniform distribution (without using a multiple restarts approach), while for the learning rate decay we use the following hyperbolic relation weighted by the square root of the current epoch
\begin{equation}
\label{learning_decay}
\eta = \frac{\eta_0}{1 + \sqrt{\text{epoch}}} .
\end{equation}
This way we reach a better accuracy, since the learning parameter gives a weighted importance to the information at any instance of the training loop on epochs.
The latter is controlled by an early-stop procedure, which checks whether or not the training validation accuracy has decreased within the last $N^{ep}_{it}$ iterations, or the loop has reached a maximum of instances $N^{ep}_{\max}$. 

Furthermore, we remark that a critical task when dealing with neural networks is the by-hand tuning of the \textit{hyper-parameters}, some of which we have already introduced (number of layers, number of neurons, learning rate, activation function).  Nonetheless, a more robust investigation of the hyper-parameter space can be done through automated machine learning (AutoML) and Bayesian optimization \cite{autoML}.

At the end of the procedure, we consider the network identified by the configuration of the weights associated to the best validation accuracy and we use it for the non-intrusive step.

Moreover, we introduce a normalization on the dataset to rescale it in the range $[0, 1]$. This is a necessary step to allow the network to learn from the different order of magnitudes of the training dataset and to improve the results.

Finally, we note that although many different sampling procedures can be adopted, we chose an equispaced or log-equispaced distribution, since they help guarantee that after the splitting of the dataset, each portion of the parameter domain is well represented for the training and validating phases, also when $N_{train}$ is small. Another observation is that we use a fixed initialization seed to guarantee the reproducibility of these results.

All ad-hoc settings for the network that one can implement increase the computational burden, so that the learning procedure may become too costly. Given the low-dimensionality of the input and output layers, we always rely on small network structures, thus avoiding more complex deep neural networks and machine learning's advanced techniques. This allow us to have a rapid training procedure to be embedded into the costly HF offline phase. 

As concerns the errors,  we consider, for a given parameter $\boldsymbol{\mu} \in \mathcal{P}$, the quantities $$\epsilon_{RB}(\boldsymbol{\mu}) = \frac{\norm{X_\N(\boldsymbol{\mu}) - X_N(\boldsymbol{\mu})}_{\X_{\N}}}{\norm{X_\N(\boldsymbol{\mu})}_{\X_{\N}}} \qquad \text{and} \qquad \epsilon_{NN}(\boldsymbol{\mu}) = \frac{\norm{X_\N(\boldsymbol{\mu}) - X_{NN}(\boldsymbol{\mu})}_{\X_{\N}}}{\norm{X_\N(\boldsymbol{\mu})}_{\X_{\N}}} ,$$ respectively for the POD-Galerkin and the POD-NN relative errors.

We evaluate these errors on the test parameter set $\Xi_{te} \subset \mathcal{P}$ of dimension $N_{te}$, which is the parameter subspace on which we build the bifurcation diagrams. As statistics for the performance, we considered the average of these quantities, and denote them by $$\overline{\epsilon}_{RB} = \frac{\sum_{\boldsymbol{\mu} \in \Xi_{te}}\epsilon_{RB}(\boldsymbol{\mu})}{N_{te}} \qquad \text{and} \qquad \overline{\epsilon}_{NN} = \frac{\sum_{\boldsymbol{\mu} \in \Xi_{te}}\epsilon_{NN}(\boldsymbol{\mu})}{N_{te}} ,$$
and their maximum value
$$\epsilon_{RB}^{\max} = \max_{\bmu \in \Xi_{te}} \epsilon_{RB}(\bmu) \qquad \text{and} \qquad \epsilon_{NN}^{\max} =\max_{\bmu \in \Xi_{te}} \epsilon_{NN}(\bmu) ,$$
over the whole test data set $\Xi_{te}$.
We use the same notation also when the unknown $X$ contains multiple variables, thus specifying the corresponding field.

Instead of considering a network which aims at approximating all coefficients monolithically, i.e.\ trying to recover simultaneously, in a CFD context, the reduced vector for velocity and pressure with the same network, we construct two different NNs with their own weights for each field. This is crucial to obtain good approximation results, indeed even if the networks share the same inputs, the corresponding outputs usually have very different behaviour and magnitude. As in \cite{hesthaven2018non}, we have not used the supremizer to recover the POD-NN solution.

Before moving to the numerical results, we present an empirical detection tool to recover the position of the bifurcation points in multi-parameter applications. We will use this technique to understand how changes in the domain geometry affect the bifurcating phenomena.

\subsection{A Reduced Manifold Based Bifurcation Diagram (RMBBD)}\label{sec:redmanbbd}
We aim to develop a non-intrusive empirical strategy, based on theoretical considerations, to recover the bifurcation diagram of  potentially unknown bifurcating phenomena from the reduced coefficients in the RB expansion. As we have discussed in previous sections, the investigation of the critical points for multi-parametrized models can be too expensive, since we need to consider outputs relying on the high-fidelity approximations. Thus, the starting point to obtain an efficient prediction about the position of the bifurcation points will be the POD-NN technique.
 
To catch the bifurcating phenomena, we usually need a priori information about the position of the critical points to refine the parameter test space and allow for a better description of the branching. We aim at efficiently reconstructing a bifurcation diagram where the output is entirely based on the reduced coefficients vectors.
This approach can be pursued with the reduced coefficients obtained by POD-NN or RB techniques, i.e.\ from the output of the neural network or the solution of the RB projection.
Nevertheless, the main issue with the pure RB approach is that we need a reasonably rich set of reduced vectors, which is infeasible in the absence of hyper-reduction technique. For this reason, all simulations are obtained from the evaluation of the network.  

The paradigm of ROMs is the approximation of the manifold of the truth solutions by means of a reduced space spanned by some suitable basis functions, here obtained through POD.
But when dealing with bifurcating models, this results in a much more difficult task, since the solutions near a critical point suddenly change their behaviour, exhibiting a $\mathcal{C}^1$ discontinuity in the chosen bifurcation output.

The main idea is to take advantage of the non-smoothness of the manifold to construct a detection tool which is able to track the bifurcation points.  
Indeed, if we are able to reconstruct the bifurcation diagram in the RB space with good accuracy, since the basis are fixed and precomputed, it means that also the reduced coefficients have to show a sudden variation, as they encode the parametrized information.


If we consider the \textit{sensitivity} of the POD-NN solution, i.e., the partial derivative $\frac{\partial X_{NN}(\bmu)}{\partial \mu_i}$, then by the orthonormality of the ROM's basis functions, we obtain 
$$\norm{\frac{\partial X_{NN}(\bmu)}{\partial \mu_i}}_{\X_{\N}} = \ \norm{\sum_{k=1}^N \frac{\partial X_{NN}^{(k)}(\bmu)}{\partial \mu_i} \Sigma^k}_{\X_{\N}} \leq \ \sum_{k=1}^N \abs{\frac{\partial X_{NN}^{(k)}(\bmu)}{\partial \mu_i}} ,$$
so that the behaviour of the solution w.r.t.\ $\bmu$, when expressed in terms of the POD basis, can be controlled by the vectors of reduced coefficients, thus giving us information about the branching.
Motivated by this consideration we choose, as a detection tool, an approximation of the curvature of the \textit{reduced coefficients manifold}. 
Assuming that the RB space contains enough information, the reduced coefficients have to reflect the bifurcation phenomenon, showing the maximum curvature w.r.t.\ the parameter space near the critical points.


Let us describe in more detail the approach for the multi-parameter case. For the sake of clarity we consider the case with $P = 2$. 

If we consider the parameter $\bmu = (\mu_1, \mu_2) \in \Pa$, we can efficiently and non-intrusively compute $\mathsf X_{NN}(\bmu) $ for each pair of the grid $G = \{(\mu_1^{(i)}, \mu_2^{(j)})\}_{(i,j)=1}^{(n,m)} \subset \Pa$. 
This way, we obtain a 3D tensor defined as $\{\mathsf X_{i,j}^k\}_{(i,j,k)=1}^{(n,m,N)}$, such that for every $\bmu \in G$ the vector $\{\mathsf X_{i,j}^k\}_{k=1}^N$ is, e.g., the reduced coefficients expansion obtained through the POD-NN technique, i.e.\ $\mathsf X_{NN}(\bmu) = \{X_{NN}^{(k)}(\bmu)\}_{k=1}^N$.


The pseudo-code in Algorithm \ref{alg:03}, illustrates the procedure to find the critical points. 
For each value of $\mu_2$, we define the curvature of the sum over all the $N$ coefficients vector $\mathsf X_{i,k}$, by means of the Matlab function $\mathsf{del2}$.
The latter, representing the divergence of the gradient, measures how much the value of the field differs from its average value taken over the surrounding points, and provides a valid measure for the actual curvature. 
Having obtained, for each point in the grid $G$, the value of the corresponding curvature,  we can simply compute the maximum value of this quantity for each fixed $\mu_2$, and obtain a curve in $G$ which describes the critical points location w.r.t.\ $\mu_1$.

\begin{algorithm}
\caption{Reduced manifold based bifurcation diagram}\label{alg:03}
\begin{algorithmic}[1]
\For{$j = 1:m$}\Comment{Loop over the parameter space}
	\State{$\mathsf X_{i,k} = \mathsf X_{i,j}^k \in \mathbb{R}^{n \times N}$}  \Comment{Extract the reduced coefficients for all $\bmu$}
	\State{$\mathsf{C}_j = \sum_{k=1}^N \Delta \mathsf X_{i,k} \in \mathbb{R}^{n}$}\Comment{Compute the global curvature}
\EndFor	
\State{$M = \argmax_i \Delta\mathsf{C}_{i,j} \in \mathbb{R}^m$}\Comment{Find indexes of local maxima for the curvature}
\State{$x^{*} = \{\mu^{(i)}_1\}_{i \in M}$}
\Comment{Extract the vector of critical points}
\end{algorithmic}
\end{algorithm}
%
%
%



\section{Numerical results}\label{sec:results}
In this section, we present two different benchmarks concerning the investigation of POD-NN approximation properties when dealing with bifurcating systems in computational fluid-dynamics problems governed by the Navier-Stokes equations. In particular, we consider multi-parameter applications with parametrized geometries linked to the Coanda effect in a sudden-expansion channel and to the lid driven flow in a triangular cavity.

\subsection{Navier-Stokes system in a channel}
\label{subsec:ns_cha}
Let us begin by describing the complex phenomena originating in the Navier-Stokes equations when modelling the so-called Coanda effect \cite{Quaini,pichistrazzullo,pintore2019efficient,HESS2019379,Hess2019,pitton_quaini_2017,pitton2017}. The latter expresses the tendency of a fluid jet to be attracted to a nearby surface. Being present in many practical applications, such as a cardiac disease related to mitral valve regurgitation, it is important to understand its behaviour. Indeed, if we consider a channel geometry, a fluid characterized by an high viscosity will present a jet which is symmetric w.r.t.\ the horizontal axis. As we lower the viscosity, the inertial effects of the fluid become more important and the two symmetric recirculation regions, formed downstream of the expansion, break the symmetry to give rise to an asymmetric jet, which represents the \textit{wall-hugging} behaviour. Furthermore, the new configuration inherits the stability properties of the symmetric configuration \cite{pichistrazzullo}, but it also represents a source of trouble and inaccurate measurements in the biomedical context.

From the mathematical point of view, the problem described above translates to a generic parametrized PDE of the form \eqref{eq:strong_form} which admits, when decreasing the viscosity $\mu$ below a certain critical threshold $\mu^*$,  the coexistence of multiple solutions for the same value of the parameter $\mu \in \mathcal{P}$.

\begin{figure}
\centering
\def\svgwidth{\linewidth}
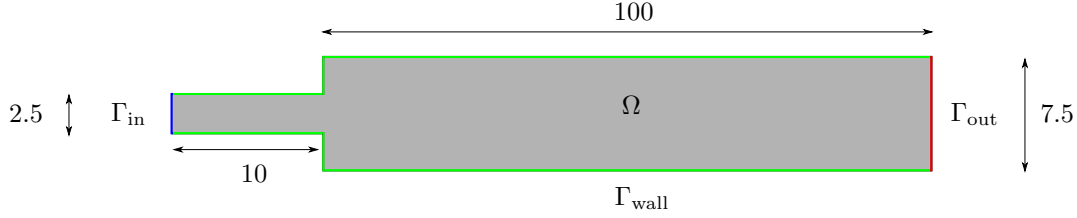
\caption{Domain $\Omega$ which represents a straight channel with a narrow inlet.}
\label{fig:channel}
\end{figure}

To be concrete, we consider a viscous, steady and incompressible flow, in the long planar straight channel with a narrow inlet represented by the domain $\Omega \subset \R^2$, depicted in Figure \ref{fig:channel}, which aims at reproducing a toy version of the heart chamber downstream the mitral valve.

The flow inside the sudden-expansion channel can be modelled by the Navier-Stokes system
\begin{equation}
\label{NS_eq}
\begin{cases}
u\cdot\nabla u -\mu \Delta  u + \nabla p=0 \quad &\text{in} \ \Omega, \\
\nabla \cdot u  =  0 \quad &\text{in} \ \Omega,
\end{cases}
\end{equation}
where $u = (u_x, u_y)$ is the velocity, $p$ the pressure normalized over a constant density and $\mu$ the viscosity.
We supplement system \eqref{NS_eq} with proper boundary conditions (BCs): a stress free BC on the velocity on $\Gamma_{\text{out}}$, a no-slip Dirichlet BC on the physical walls $\Gamma_{\text{wall}}$ and a parabolic Dirichlet BC at the inlet $\Gamma_{\text{in}}$, as follows
\begin{equation*}
\begin{cases}
u = u_{\text{in}}  \quad &\text{on} \ \Gamma_{\text{in}}, \\
u = 0  \quad &\text{on} \ \Gamma_{\text{wall}}, \\
- pn  + (\mu \nabla u) n  = 0  \quad &\text{on} \ \Gamma_{\text{out}},
\end{cases} \quad \text{where} \quad
 u_{\text{in}} = \begin{bmatrix} 20(5-y)(y-2.5) \\ 0 \end{bmatrix}.
\end{equation*}
Furthermore, we denote the function spaces $\V_{\text{in}}=\{v \in \left(H^1(\Omega)\right)^2 \mid v=v_{\text{in}} \text{ on }\Gamma_{\text{in}}, \ v=0 \text{ on }\Gamma_{\text{wall}}\}$, $\V_0=\{v \in \left(H^1(\Omega)\right)^2 \mid v=0 \text{ on }\Gamma_{\text{in}} \cup \Gamma_{\text{wall}}\}$ and $\Q=L^2(\Omega)$ respectively for the velocity (keeping track of the inhomogeneous Dirichlet BC) and the pressure fields. The variational formulation reads as: given $\mu \in \mathcal{P}$, find $ u \in \V_{\text{in}}$ and $p \in \Q$ such that
\begin{equation}
\label{eq:gal_ns}
\left\{
\begin{aligned}
 \mu\int_\Omega\nabla  u\cdot\nabla  v\ d\Omega +\int_\Omega \left( u\cdot\nabla u\right) v\ d\Omega - \int_\Omega p\nabla\cdot  v\ d\Omega = 0 \quad\quad &\forall\ v \in \V_0, \\
\int_\Omega q\nabla\cdot  u\ d\Omega = 0\quad\quad &\forall\ q \in \Q.
\end{aligned}
\right.
\end{equation}
If we introduce the following bilinear and trilinear forms
\begin{equation}
\label{eq:forms}
\begin{aligned}
a(u, v; \mu) =\mu\int_\Omega\nabla u\cdot\nabla v \, d\Omega &\hspace{1cm}&&\forall \, u, v \in \V_0,\\
b(u, p) = -\int_\Omega(\nabla\cdot u) \hspace{.05cm}p \, d\Omega &\hspace{1cm}&&\forall \, u \in \V_0,\ \forall \, p \in \Q,\\
c(u, \bar u, v)=\int_\Omega \left(u\cdot\nabla \bar u\right) v \, d\Omega &\hspace{1cm}&& \forall \, u, \bar u, v\in \V_0 ,
\end{aligned}
\end{equation}
we can rewrite the formulation of \eqref{eq:gal_ns} in an equivalent way
as: given $\mu \in \mathcal{P}$, find $u \in \V_{\text{in}}$ and $p \in \Q$ such that
\begin{equation}
\label{eq:gal_ns2}
\begin{cases}
a(u,v; \mu) +c(u,u,v) +b(v,p) = 0 \quad &\forall \, v \in \V_0, \\
b(u,q) = 0\quad &\forall \, q \in \Q .
\end{cases}
\end{equation}
After a description of the physical phenomenon and its modelling equations, we can move to the results of the POD-NN approximation. In particular, we discuss the bifurcating regime starting from a one-dimensional parameter space and then investigate how this is influenced by a parametrized domain in a multi-parameter setting.



\subsubsection{The bifurcating regime}

Let us first consider the Navier-Stokes equations in the bifurcating regime, when the solution to the parametrized PDE exists but it is not unique.
During the study of the solution's behaviour varying the viscosity values, we expect the system to undergo two qualitatively different configurations: (i) a physically unstable configuration represented by a symmetric jet flow, and (ii) a physically stable configuration represented by a wall-hugging jet.

These solutions, coexisting when the viscosity is lower than a critical threshold, belong to different branches that intersect in the bifurcation point $\mu^*$ forming a \textit{pitchfork bifurcation}. Indeed, due to the symmetry group of the PDE, there exist two stable asymmetric configurations w.r.t.\ the channel's centreline.
We are only interested in the approximation of one of these two, e.g., the downwards wall-hugging behaviour. Otherwise, the reconstruction of the full bifurcation diagram would require the network to assign different outputs to the same input parameter. To solve this issue, one could exploit a branch-wise strategy, building a specific network for each branch or consider more involved machine learning strategies.
Furthermore, it is clear that a key aspect for the reconstruction of a bifurcating behaviour, especially in an artificial neural network context, is the complete identification of the singularities of the model, which can be pursued following the strategy discussed in Section \ref{sec:redmanbbd}.



For this reason, we consider as a range for the viscosity, the parameter space $\Pa = [0.5, 2]$ which includes the first bifurcation point for the model at $\mu^{*} \approx 0.96$ \cite{pichistrazzullo,pintore2019efficient,pichi_phd}.

\begin{figure}
\centering
\begin{minipage}{0.49\textwidth}
\includegraphics[width=8cm]{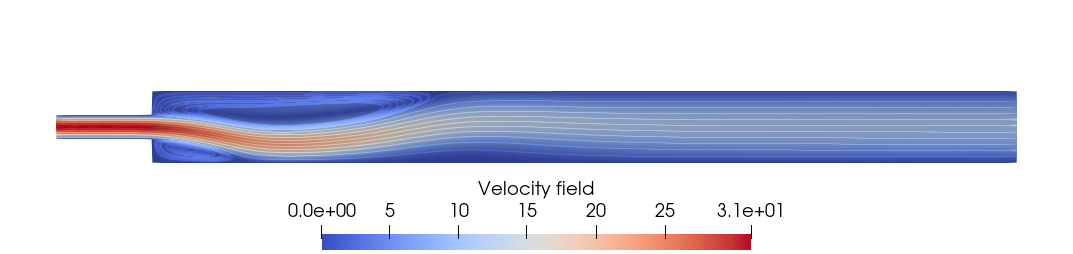}
\includegraphics[width=8cm]{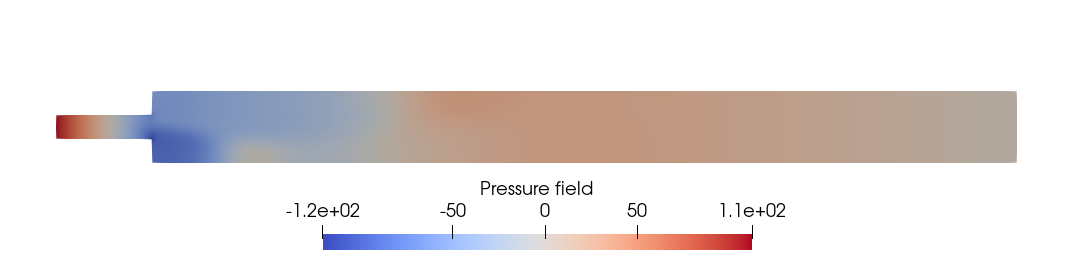}
\end{minipage}\hfill
\begin{minipage}{0.49\textwidth}
\includegraphics[width=8cm]{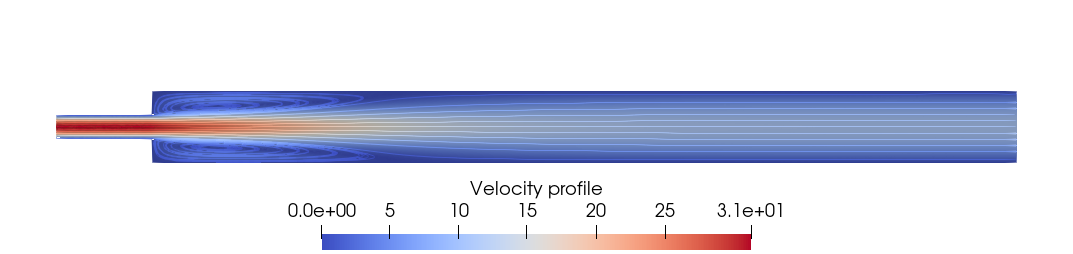}
\includegraphics[width=8cm]{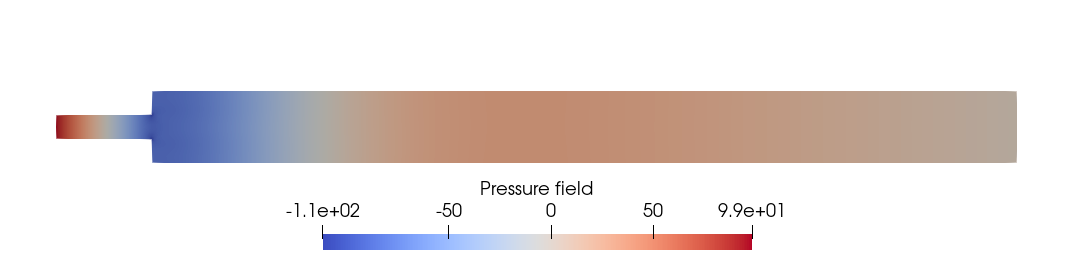}
\end{minipage}
\caption{High-fidelity post-bifurcated solutions for the Coanda model at $\mu = 0.5$, velocity and pressure fields (top and bottom), for the asymmetric and symmetric profiles (left and right).}
\label{fig:NS_sol_hf_bif_ann}
\end{figure}

We want to describe and compare the different strategies used for the approximation of the Coanda effect.
We start by showing, in Figure \ref{fig:NS_sol_hf_bif_ann}, representative FE solutions of the bifurcated states for $\mu = 0.5$, in which the velocity flow attaches to the bottom of the channel and remains symmetric, left and right respectively.

Since we are interested in the reconstruction of the bifurcation diagram, we fix the testing set $\Xi_{te}$ as an equispaced set of $N_{te} = 151$ points in $\mathcal{P}$.
This way, we are able to perform the continuation procedure, needed for the FE and RB approximations, to detect the bifurcating behaviour by following the evolution of the manifold through the choice of the initial guesses.
Once again, this is not required by the neural network, which simply evaluates the input parameters giving back the value of the reduced coefficients.

To capture all effect of the nonlinearities, which give rise to the bifurcations, we consider a fine mesh discretization resulting in
$\N = 79868$ degrees of freedom for the FE space. Regarding the RB step, by selecting the POD tolerance as $10^{-8}$, we obtain $N_u = 11$ and $N_p=9$ basis functions for the velocity and pressure fields, respectively.
We remark that the dimension of the RB space for the velocity is increased by one, due to the lifting coming from the inhomogeneous Dirichlet BC.

As concerns the learning procedure, we considered $\eta_0 = 0.3$ for both velocity and pressure networks, and we updated it by means of the learning rate schedule in \eqref{learning_decay}, considering $N^{ep}_{it} = 5 \cdot 10^2$ and $N^{ep}_{\max} = 10^4$.

The offline phase is initialized with $N_{train} = n_b^2$ (and corresponding size $n_b$ for the mini-batches, where $n_b =6(i+1)$ with $i=1,2,3$) equispaced points in the parameter set $\Pa$, using the ratio described before to split the dataset in training and validation sets.

\begin{figure}
\begin{minipage}{0.45\textwidth}
\centering
\includegraphics[width=6.5cm]{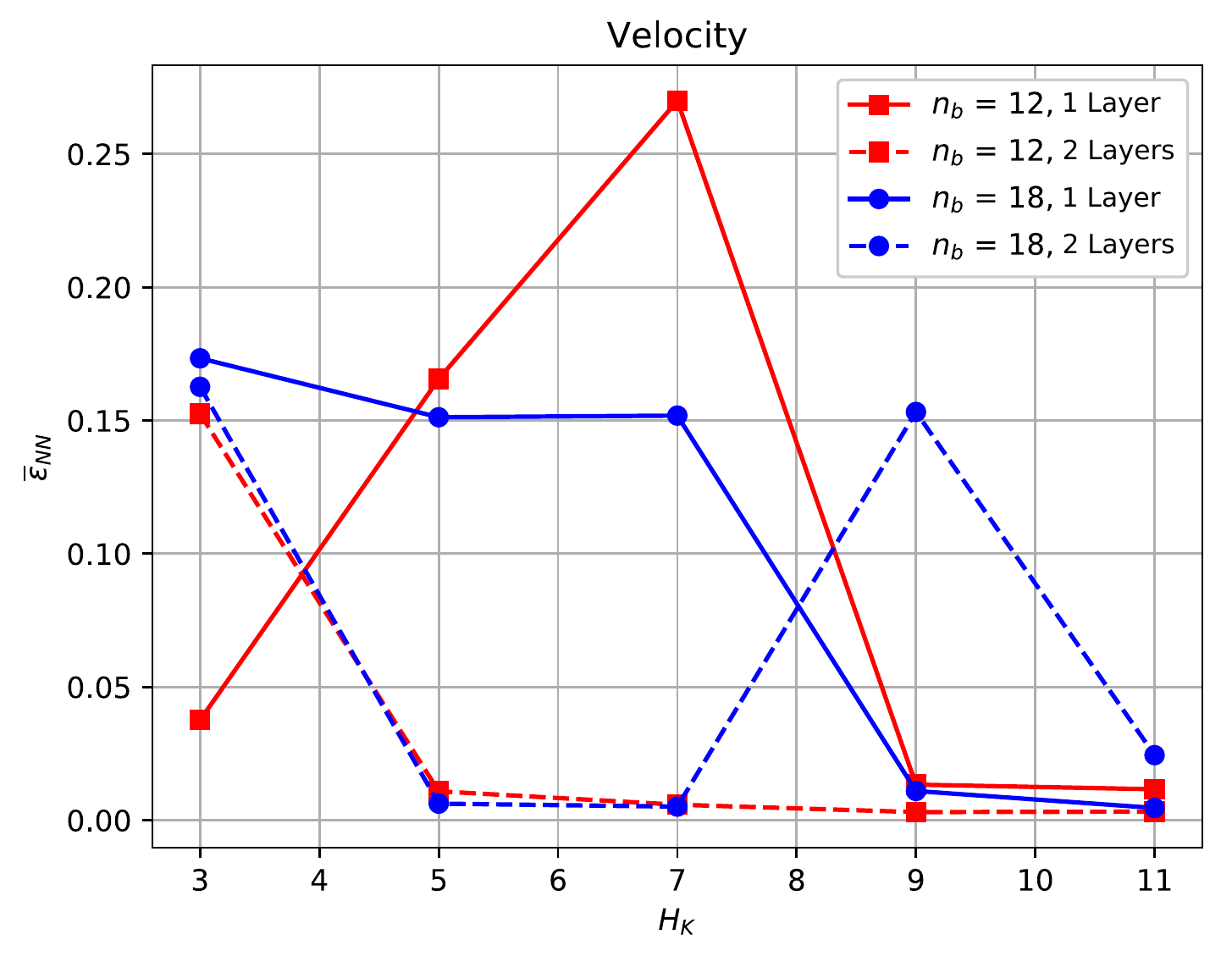}
\label{fig:err_NS_bif_v}
\end{minipage}
\begin{minipage}{0.45\textwidth}
\centering
\includegraphics[width=6.5cm]{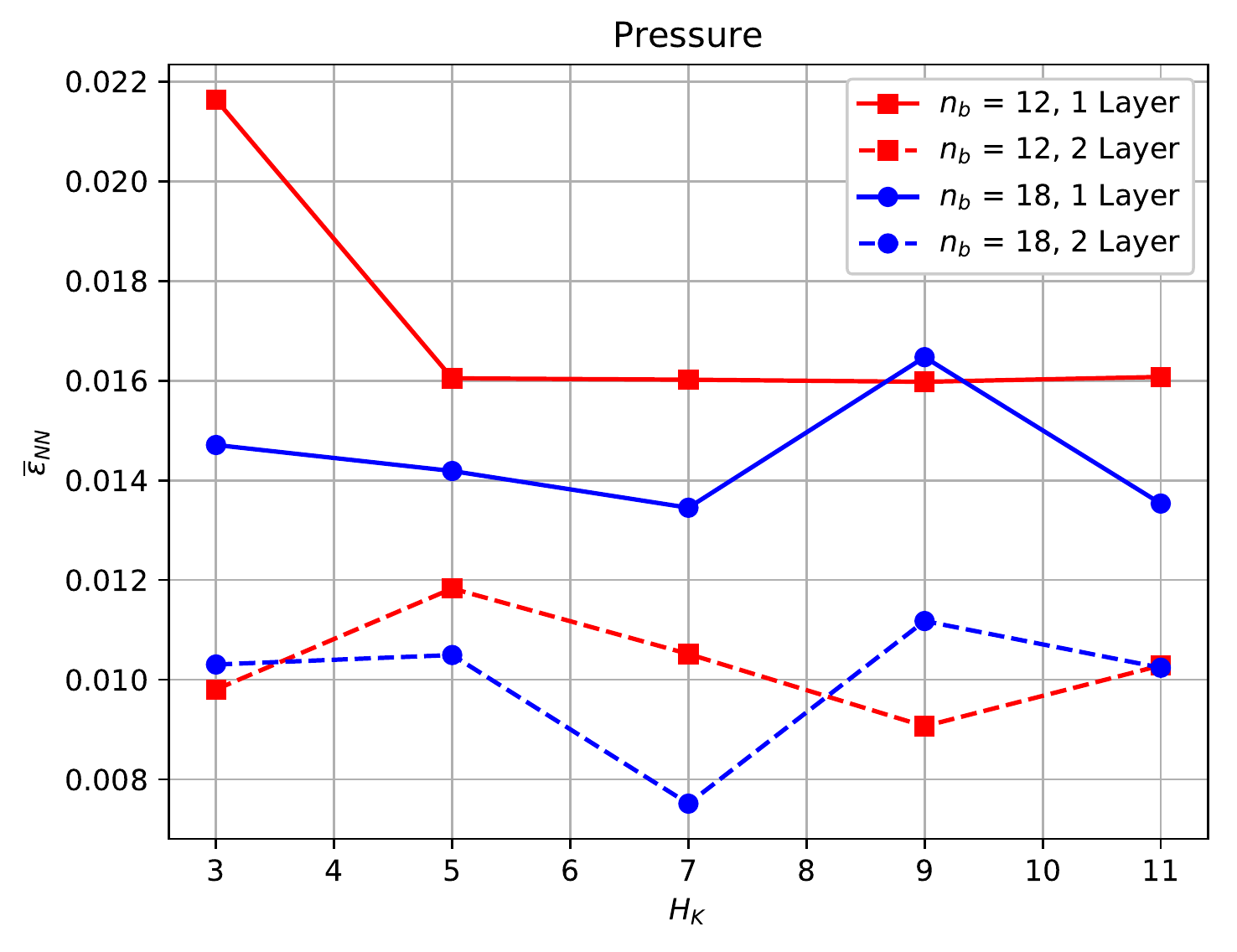}
\label{fig:err_NS_bif_p}
\end{minipage}
\caption{Mean NN errors $\overline{\epsilon}_{NN}$  w.r.t.\ the number of neurons $H_K$, for velocity and pressure fields, left and right, respectively.}
\label{fig:err_NS_bif}
\end{figure}

To find the best configuration for the network, we test the accuracy properties on $\Xi_{te}$ w.r.t.\ the dimension of the dataset, while increasing the number of hidden layers $L_K = 1,2$ and computing neurons $H_K = 3, \dots, 11$, for $n_b = 12, 18$.
In Figure \ref{fig:err_NS_bif} we show a comparison of the mean NN errors $\overline{\epsilon}_{NN}$ for the velocity and pressure fields w.r.t.\ $H_K$.

\begin{figure}[b]
\begin{minipage}{0.45\textwidth}
\centering
\includegraphics[width=6.5cm]{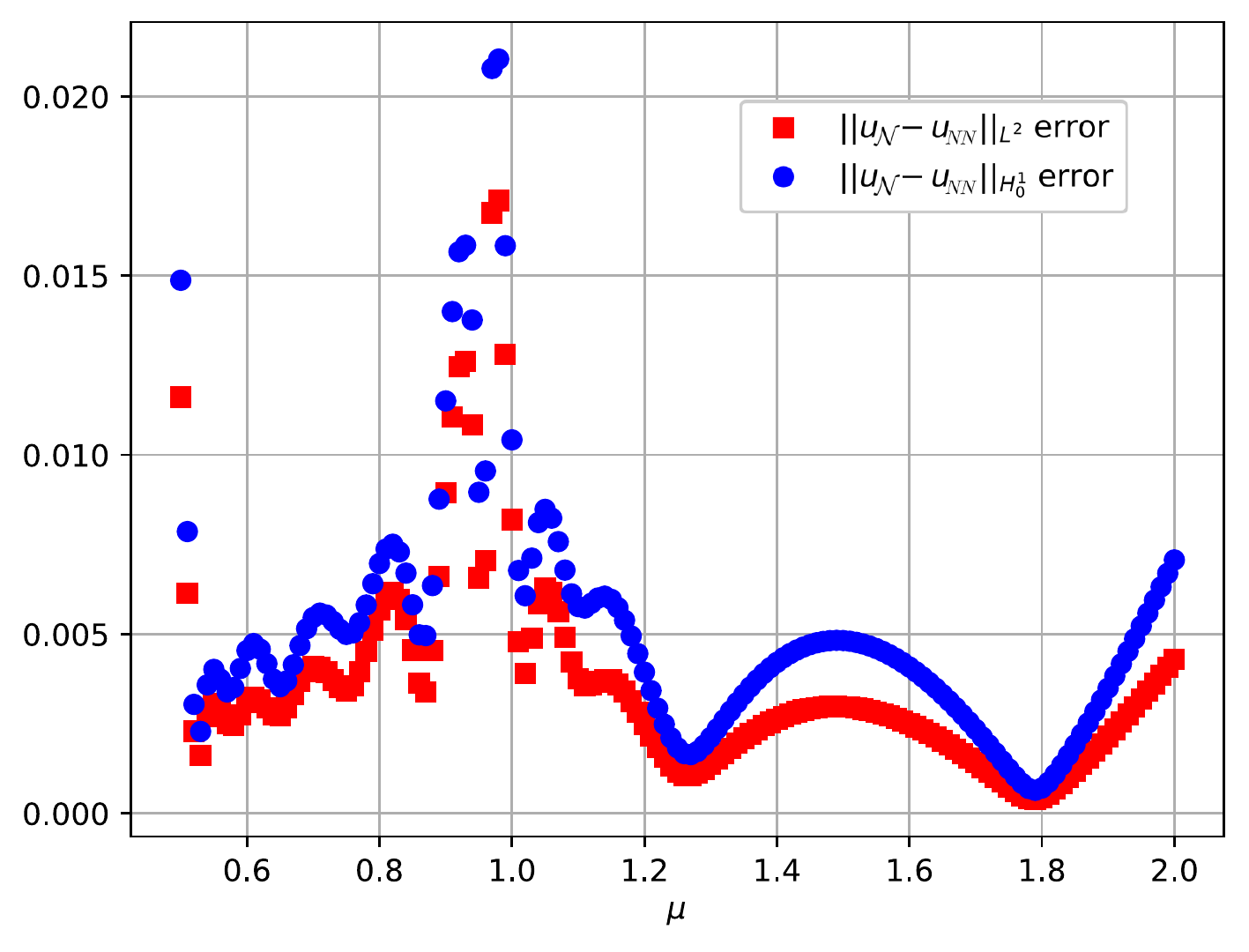}
\end{minipage}\quad\quad
\begin{minipage}{0.45\textwidth}
\centering
\includegraphics[width=6.5cm]{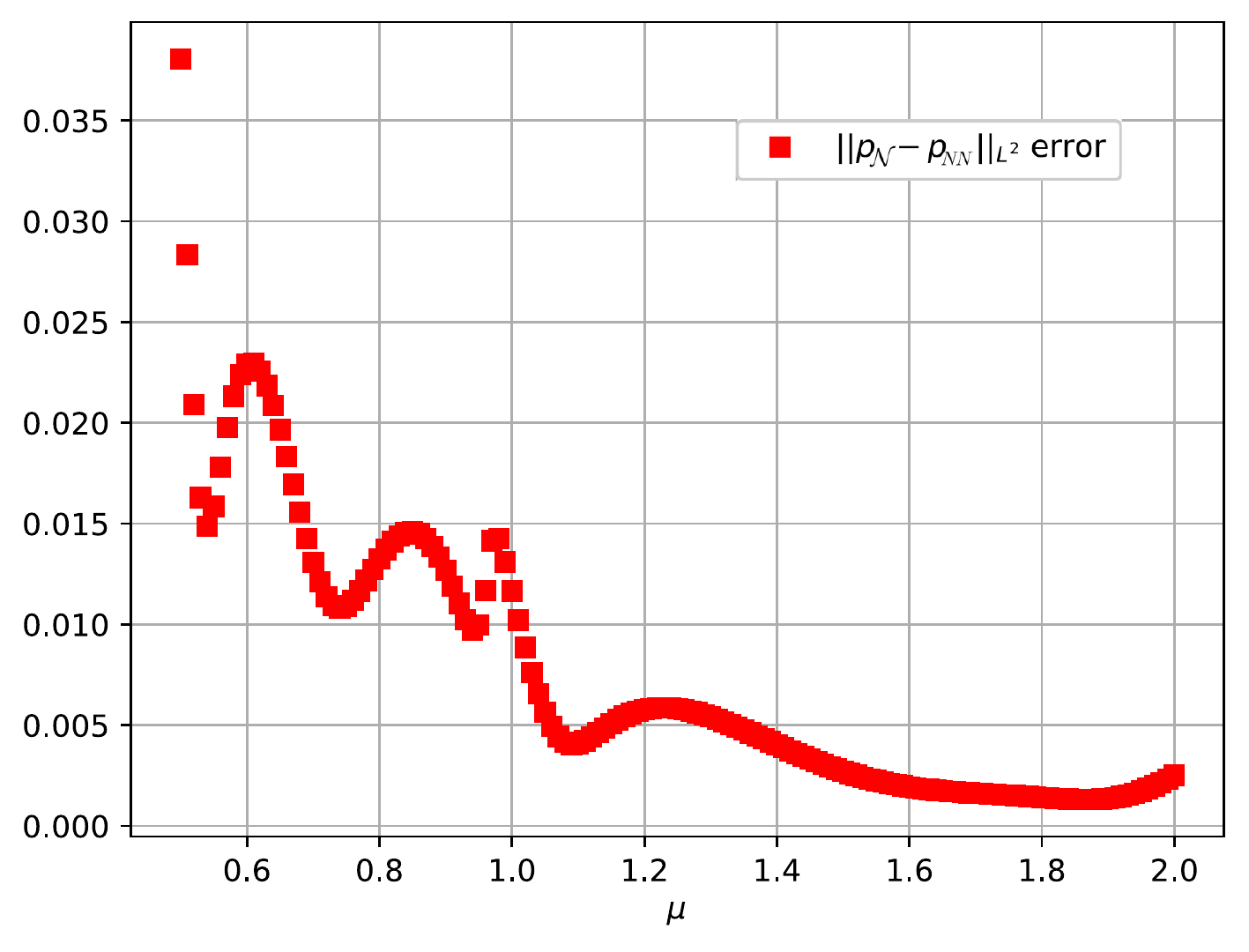}
\end{minipage}
\caption{Relative error $\epsilon_{NN}(\boldsymbol{\mu})$ for velocity and pressure fields in the bifurcating regime computed on $\Xi_{te}$, left and right, respectively.}
\label{fig:err_NS_bif_2}
\end{figure}

We show in Figure \ref{fig:err_NS_bif_2} the relative error $\epsilon_{NN}(\mu)$ for both velocity and pressure fields in the best case scenario for the network configuration, which corresponds to the case with $L_K = 2$ layers, $n_b = 18$ and $H_K = 7$. In this case, the maximum and mean errors over the testing dataset are $\epsilon^{\max}_{NN} = 0.02104$ and $\overline{\epsilon}_{NN} = 0.00506$, respectively. Similar results hold for the pressure field, where we have $\epsilon^{\max}_{NN} = 0.03805$ and $\overline{\epsilon}_{NN} = 0.00751$.

\begin{figure}
\begin{minipage}{0.45\textwidth}
\centering
\includegraphics[width=6.5cm]{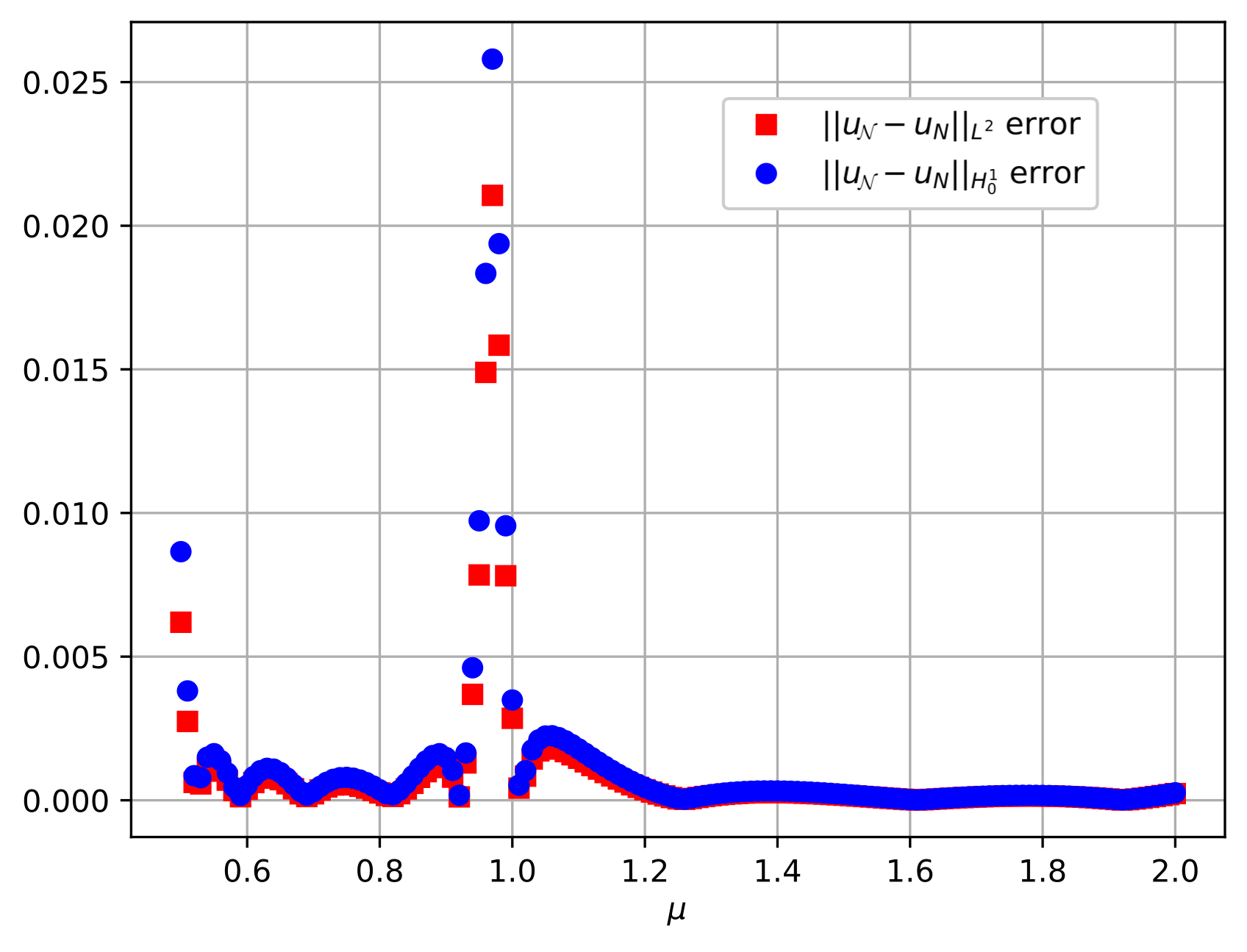}
\end{minipage}\quad\quad
\begin{minipage}{0.45\textwidth}
\centering
\includegraphics[width=6.5cm]{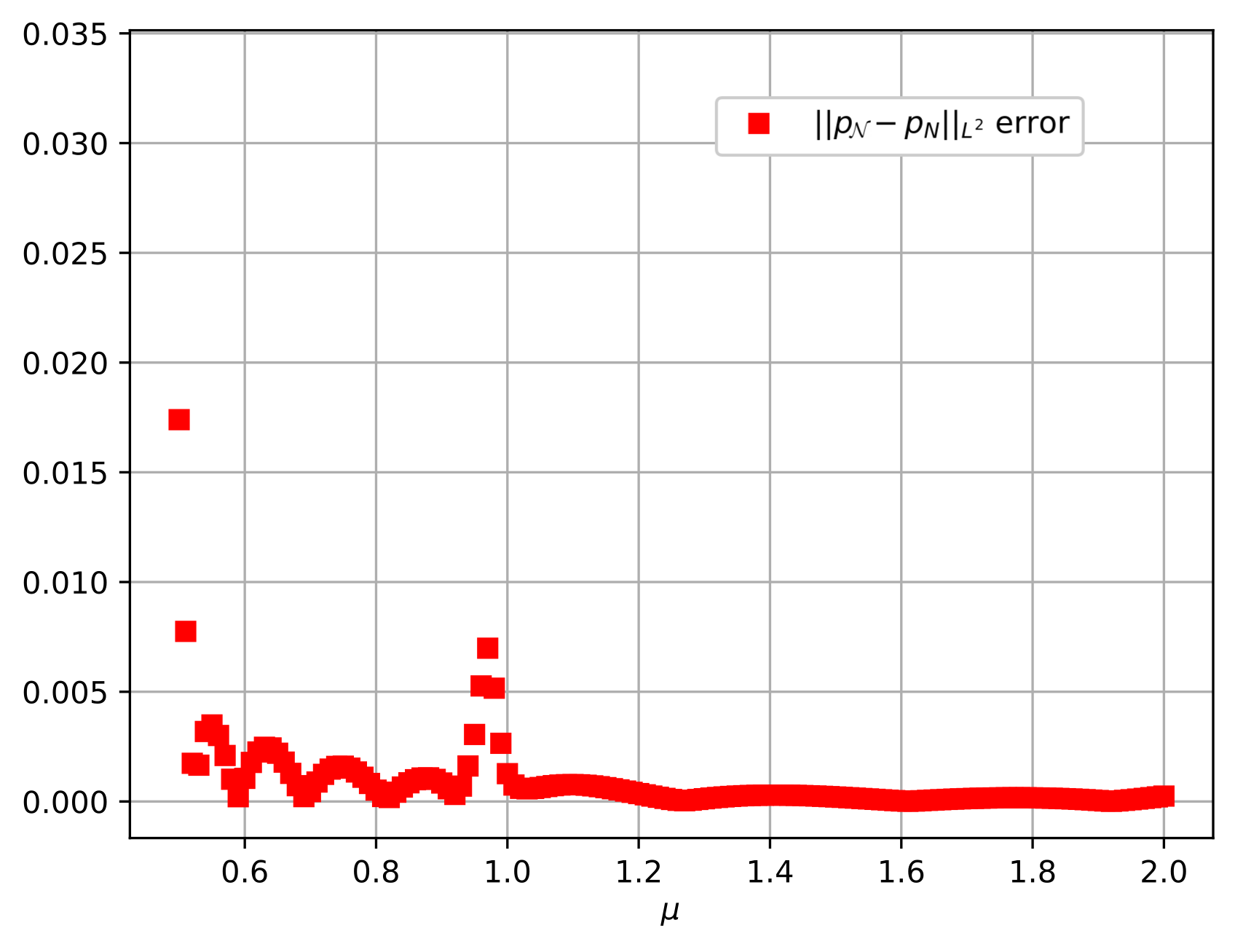}
\end{minipage}
\caption{Relative error $\epsilon_{RB}(\boldsymbol{\mu})$ for velocity and pressure fields in the bifurcating regime computed on $\Xi_{te}$, left and right, respectively.}
\label{fig:err_NS_bif_2_rb}
\end{figure}

As a comparison, we show in Figure \ref{fig:err_NS_bif_2_rb} the RB error on the testing set $\Xi_{te}$, when employing the same number of basis functions. Performing the online phase through the RB method (without hyper-reduction) we obtained the maximum and mean errors over $\Xi_{te}$ as $\epsilon^{\max}_{RB} = 0.02579$ and $\overline{\epsilon}_{RB} = 0.00115$. Again, similar results hold for the pressure field, where we have $\epsilon^{\max}_{RB} = 0.01726$ and $\overline{\epsilon}_{RB} = 0.00086$.

As seen from the errors, in the proposed setting, the POD-NN approach was able to properly recover the bifurcating behaviour of the model. Moreover, it reached good accuracy also in comparison with the RB strategy. As understood from the analysis of bifurcating problems, when the solution does not depend smoothly on the parameter, we can expect peaks in the errors corresponding to the parameter values at which the bifurcation occurs.

The results for the reduced approximation of the bifurcation test case are always less accurate when the critical point is in the parameter range.
In general it could be a limitation, but we remark that, on the contrary, since the network is only built from the snapshots, it does not see the singularities of the model and it is only slightly affected by it. This results in the similar order of accuracy between mean and max errors for the POD-NN, while the same quantities for the RB ones are of different orders of magnitude.

\begin{figure}
\centering
\includegraphics[width=8cm]{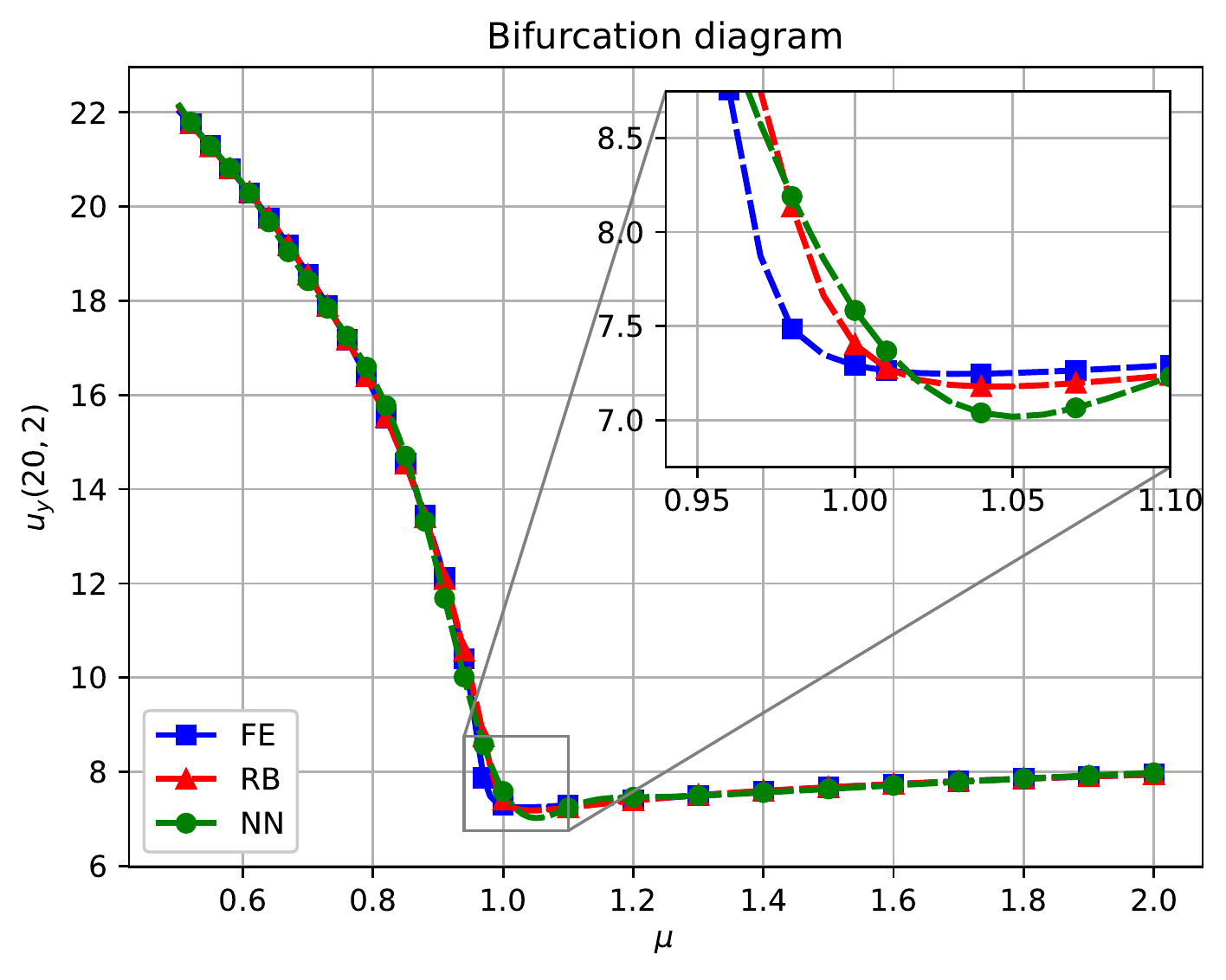}
\caption{Bifurcation diagram for the Coanda model with point-wise vertical velocity $u_y$ at $(20, 2.5)$ as output, obtained with FE, RB and POD-NN.}
\label{fig:bifurcation_ns_ann}
\end{figure}

In Figure \ref{fig:bifurcation_ns_ann} we show a comparison of the bifurcation diagrams obtained through the FE, RB and POD-NN approaches for the Coanda effect where, for each value of the viscosity $\mu$, we consider the point-wise vertical velocity $s(u) = u_y(20, 2.5)$ as output. It is clear that $\mu^*=0.96$ corresponds to the critical value at which the flow exhibits the wall-hugging behaviour.

Finally, the speed-up for the evaluation of the network on $\Xi_{te}$ in the POD-NN context, with respect to the FE method, is almost $1.2\cdot 10^6$, while the RB one is equal to $1.25$. In fact, the nonlinearity of the problem together with the lack of an efficient hyper-reduction/tensor assembly for the weak form, causes the standard RB approach to be dominated by a $\N$-dependent online phase.


%
%


\subsubsection{Multi-parameter analysis with varying geometry}\label{sec:ann_multiparam}
Having recovered the evolution of a single bifurcating branch, we move to the more complex and interesting multi-parameter case. In such case, we aim at investigating how the uniqueness regime of the system changes while varying the  model's features. Here, we generalize the Coanda's problem discussed above, by considering a geometric parametrization.
In particular, still relying on the planar sudden-expansion channel, we introduce a parametrization of the semi-width of its narrow inlet by means of the geometric parameter $w \in [0.5, 2.0]$, obtaining a problem parametrized by $\bmu = (\mu, w)$.

The context from the physical point of view is clear. By reducing or increasing the height of the inlet channel, we are changing the intrinsic value of the \textit{Reynolds number} $Re = Uw/\mu$ (where $U$ is the characteristic velocity taken at the inlet), balancing differently the inertial and viscous forces. This means that the critical points of the model can be expected to vary for the different geometries considered. More specifically, we expect to observe that the smaller the parameter $w$, and thus the corresponding inlet, the sooner the bifurcation occurs. As we will see, such behaviour will be not entirely obvious.

Let us consider the Navier-Stokes system in \eqref{NS_eq}, but this time defined over the parametrized domain $\Omega(w)$. We consider affine transformation maps of the form $\boldsymbol{\Phi}(\textbf{x}; w) = \mathbb{B}(w)\textbf{x} + c(w)$, choosing as reference domain $\overline{\Omega} = \Omega(w = 1.25)  = \bigcup_{i=1}^3 \Omega_i$, where $\Omega_i$ are defined as the three horizontal stripe-shaped subdomains, i.e.\ $\Omega_1 = [10, 100] \times [5, 7.5]$, $\Omega_2 = [0, 100] \times [2.5, 5]$ and $\Omega_3 = [10, 100] \times [0, 2.5]$. In this case, the transformation maps are given by the following quantities
\begin{alignat*}{3}
&\mathbb{B}_1(w) = \begin{bmatrix} 1 & 0 \\ 0 & 1.5 - 0.4 w\end{bmatrix} \quad & &\text{and} \quad &&c_1(w) = \begin{bmatrix} 0  \\ 3w - 3.75  \end{bmatrix}, \\
&\mathbb{B}_2(w) = \begin{bmatrix} 1 & 0 \\ 0 & 0.8 w\end{bmatrix} \quad & &\text{and} \quad &&c_2(w) = \begin{bmatrix} 0  \\ -3w + 3.75  \end{bmatrix}, \\
&\mathbb{B}_3(w) = \begin{bmatrix} 1 & 0 \\ 0 & 1.5 - 0.4 w\end{bmatrix} \quad & &\text{and} \quad &&c_3(w) = \begin{bmatrix} 0  \\ 0  \end{bmatrix}.
\end{alignat*}
Thanks to the change of variable formula \cite{QuarteroniManzoniNegri2015}, if we express with $\mathbb{J}_{\boldsymbol{\Phi}} = \mathbb{B}$ the Jacobian of the affine maps, the variational forms in \eqref{eq:forms} become

\begin{equation}
\label{eq:forms_param}
\begin{aligned}
a(u, v; \bmu) = \mu \sum_{i,j = 1}^2 \int_\Omega\frac{\partial v}{\partial x_i} \left[\mathbb{J}_{\boldsymbol{\Phi}}^{-1}(\textbf{x}; w) \mathbb{J}_{\boldsymbol{\Phi}}^{-T}(\textbf{x}; w) |\mathbb{J}_{\boldsymbol{\Phi}}(\textbf{x}; w)|\right]_{i,j} \frac{\partial u}{\partial x_j}\, d\Omega &\hspace{0.5cm}&&\forall \, u, v \in \V_0,\\
b(u, p; \bmu) = -\sum_{i,j = 1}^2 \int_\Omega p \left[\mathbb{J}_{\boldsymbol{\Phi}}^{-T}(\textbf{x}; w) |\mathbb{J}_{\boldsymbol{\Phi}}(\textbf{x}; w)|\right]_{i,j} \frac{\partial u_i}{\partial x_j}\, d\Omega &\hspace{0.5cm}&&\forall \, u \in \V_0,\ \forall \, p \in \Q,\\
c(u, \bar u, v; \bmu)= \sum_{i,j,k = 1}^2 \int_\Omega u_i \left[\mathbb{J}_{\boldsymbol{\Phi}}^{-T}(\textbf{x}; w) |\mathbb{J}_{\boldsymbol{\Phi}}(\textbf{x}; w)|\right]_{j,i} \frac{\partial \bar u_k}{\partial x_j} v_k\, d\Omega &\hspace{0.5cm}&& \forall \, u, \bar u, v\in \V_0 .
\end{aligned}
\end{equation}
Finally, we write the weak formulation of the problem, posed over the reference domain $\overline{\Omega}$, as

\begin{equation}
\label{eq:gal_ns_param}
\begin{cases}
\begin{split}
a(u,v; \bmu) + c(u,r_{in},v; \bmu) &+ c(r_{in},u,v; \bmu) + c(u,u,v; \bmu) \\&+ b(v,p;\bmu) = - a(r_{in},v; \bmu)  - c(r_{in},r_{in},v; \bmu) \end{split} \quad &\forall \, v \in \V_0, \\
b(u,q,\bmu) = -b(r_{in}, q; \bmu) \quad &\forall \, q \in \Q ,
\end{cases}
\end{equation}
where we already encode the lifting function $r_{in}$ s.t.\ $r_{in}|_{\Gamma_{in}} = u_{in}$. The solution is given by $u + r_{in}$.

\begin{figure}[t]
\centering
\includegraphics[width=14cm]{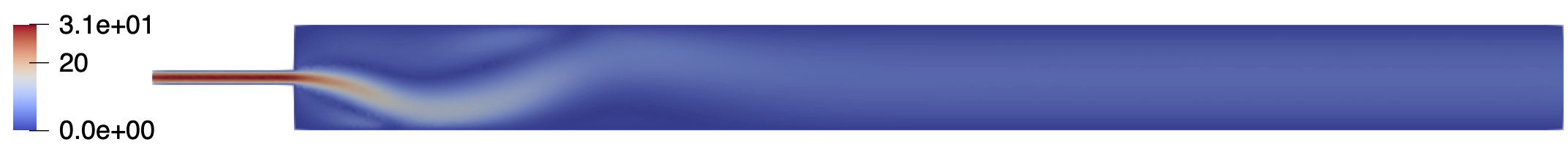}
\includegraphics[width=14cm]{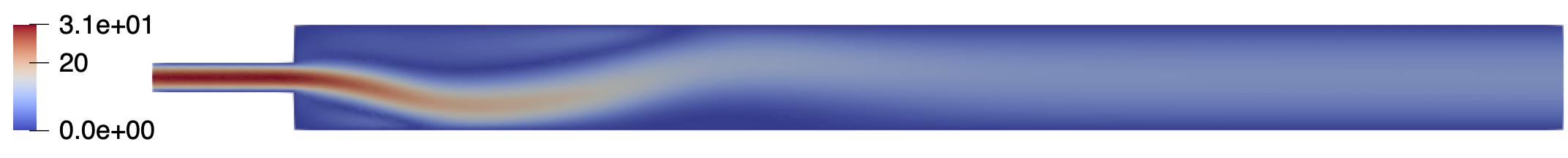}
\includegraphics[width=14cm]{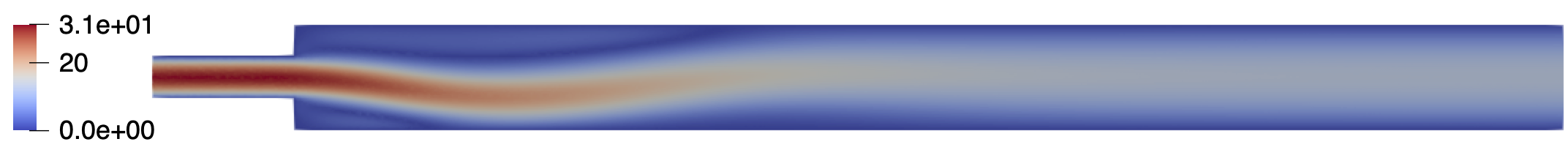}
\includegraphics[width=14cm]{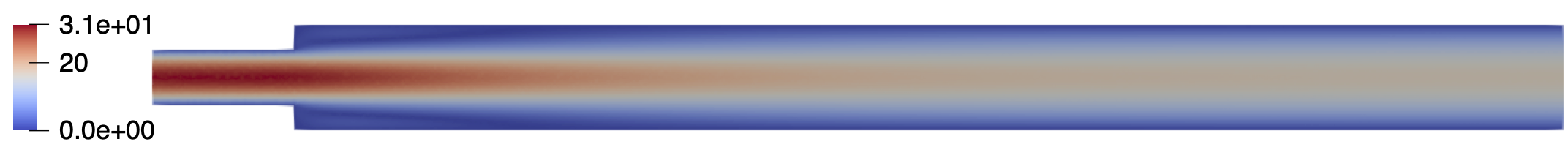}
\caption{High-fidelity velocity profiles at $\mu =0.5$ for the Coanda model with $w = 0.5, 1.0, 1.5, 2.0$, from top to bottom.}
\label{fig:NS_sol_hf_param}
\end{figure}

As expected, the new system is characterized by different regimes for its well posedness, depending on the configuration we are interested in.
As an example, we show in Figure \ref{fig:NS_sol_hf_param} four FE solutions at $\mu = 0.5$ for different values of the inlet width $w$. As we see, the system ranges from exhibiting a fully developed wall-hugging behaviour to still undergoing the pre-bifurcation regime.

These velocity profiles are already telling us that, while varying the geometry, the position of the bifurcation points changed for each fixed value of $w$. Thus, if we are able to correctly explore the parameter space during the offline phase, we can efficiently reconstruct and investigate the 3-D bifurcation diagram during the online phase, either by means of an RB or a POD-NN strategy.
As we observe empirically from the velocity profiles in Figure \ref{fig:NS_sol_hf_param}, the bifurcation phenomena happen for values of $w$ in $[0.5,1.5]$. Thus, we restrict the study to this range for the geometrical parameter.

\begin{figure}
\centering
\includegraphics[height=5.8cm]{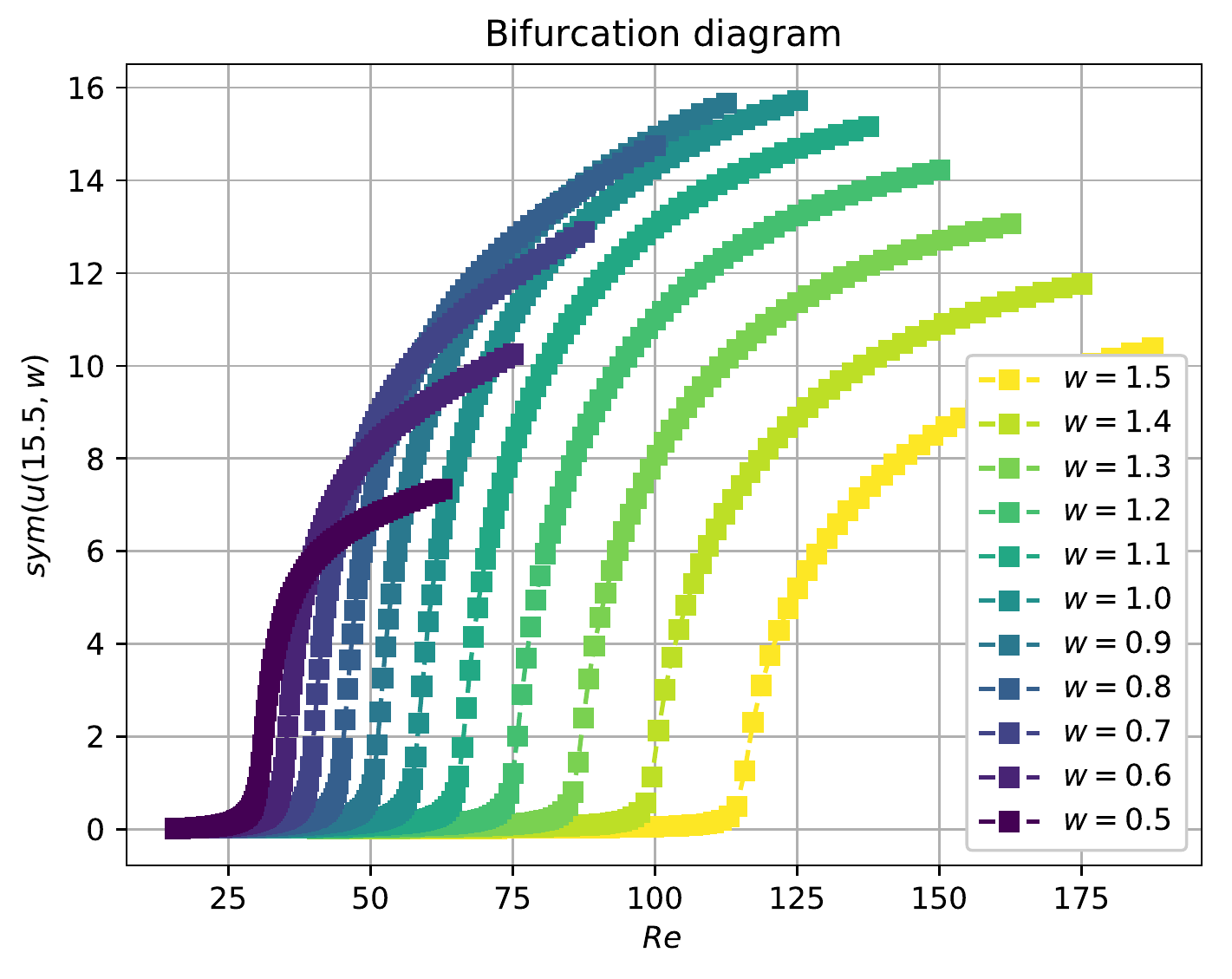}\hfill
\includegraphics[height=5.8cm]{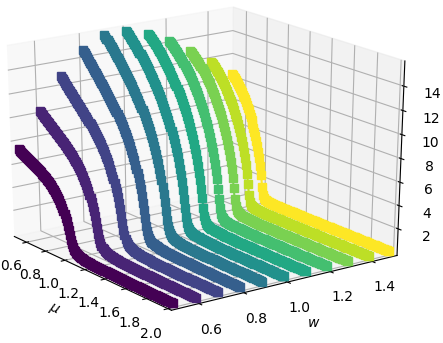}
\caption{2-D and 3-D bifurcation diagrams for the Coanda model with geometrical parametrization of the narrow inlet, with respect to the Reynolds number $Re$ and the viscosity $\mu$, left and right, respectively.}
\label{fig:bifurcation_param}
\end{figure}

We present the bifurcation diagrams in Figures \ref{fig:bifurcation_param}, respectively a 3-D version with respect to the viscosity $\mu$ and the 2-D one with respect to the Reynolds number corresponding to the chosen geometry. For the output, we chose a measure of the symmetry of the flow w.r.t.\ the mid-line of the channel. The effect of the geometrical parametrization on the location of the critical points is now evident.

We note that we are still considering an output which depends on the $\N$ degrees of freedom of the high-fidelity solution. In fact, we need to project the RB solution over the FE space to obtain point-wise evaluation of the fields. Different affinely decomposable functionals can be considered, but in this context we rely on the strategy developed in Section \ref{sec:redmanbbd} to fully exploit the artificial neural network capabilities, efficiently recovering the bifurcation diagram during the online phase, without involving any $\N$-dependent quantity.

As concerns the offline sampling procedure, the parameter space considered for this test case is $(\mu, w) \in \Pa = [0.5, 2.0] \times [0.5, 1.5]$, where we chose an equispaced grid of  $n_b^2 \times 6$ values for the viscosity and inlet's width, respectively.

Since the reduced model in this multi-parameter context has to encode much more information, a POD tolerance equal to $\epsilon_{POD} =10^{-8}$ forced us to consider $N_u = 50$ velocity basis functions and  $N_p = 24$ pressure basis functions to built the reduced manifold.

\begin{figure}
\begin{minipage}{0.45\textwidth}
\centering
\includegraphics[width=6.5cm]{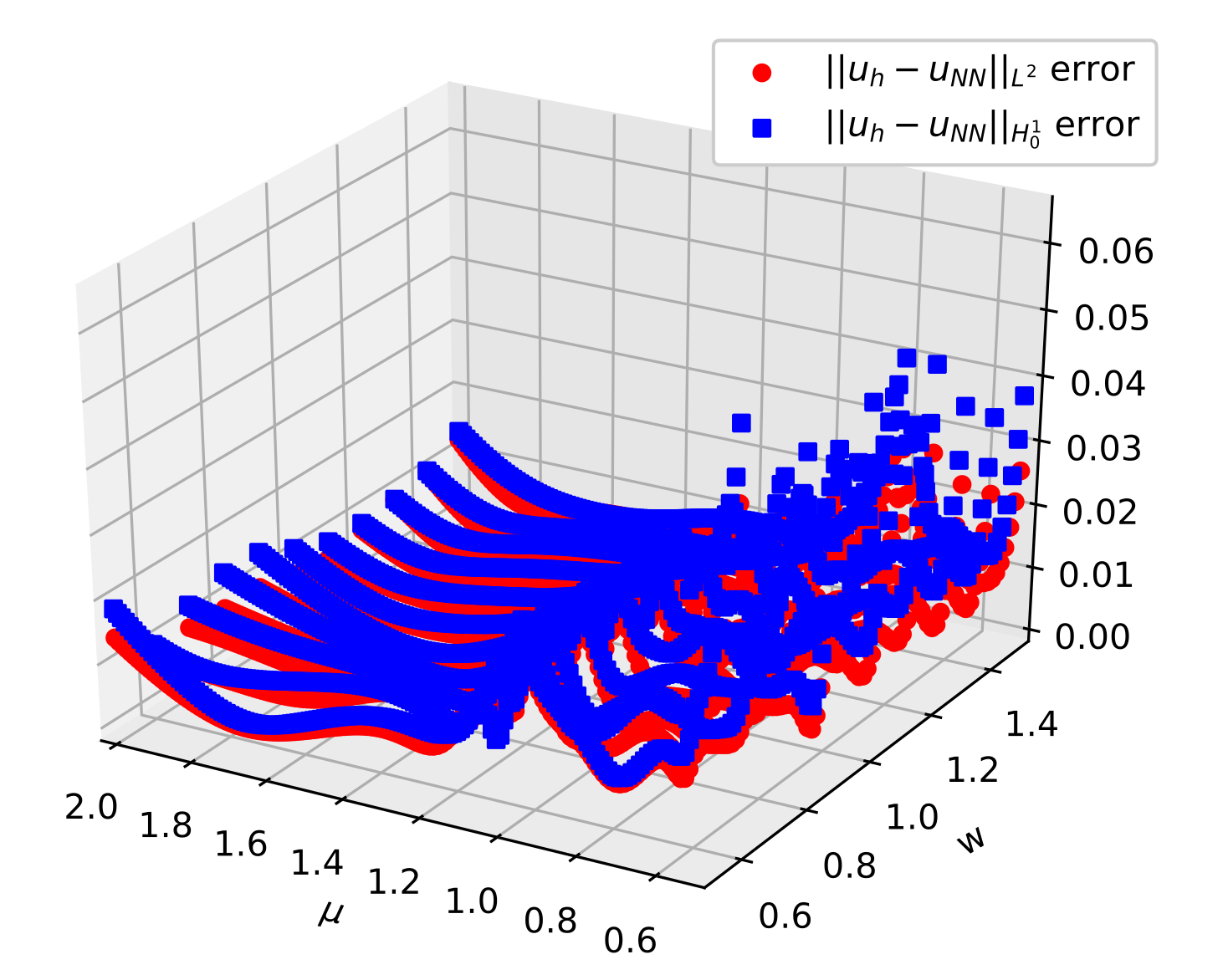}
\end{minipage}
\begin{minipage}{0.45\textwidth}
\centering
\includegraphics[width=6.5cm]{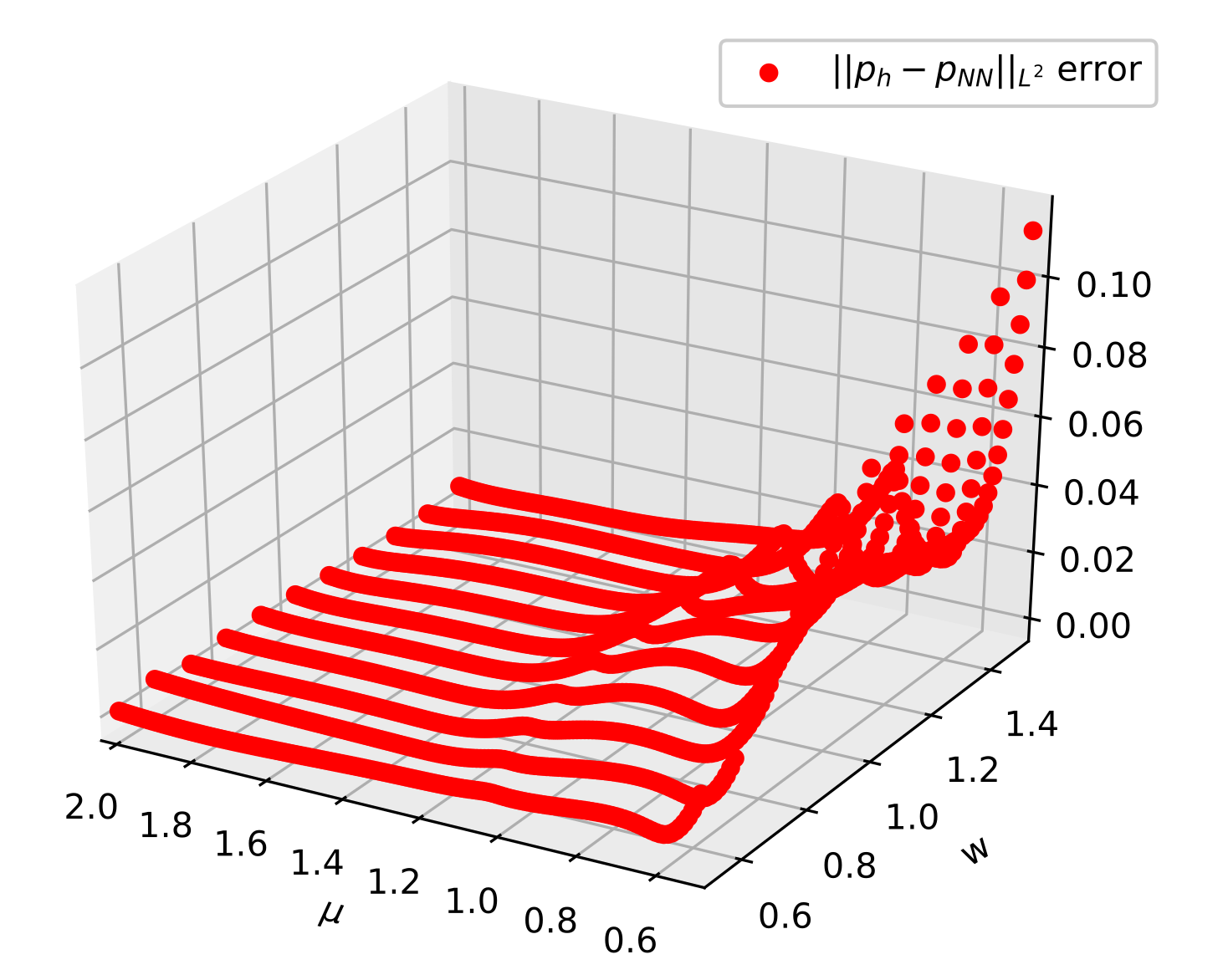}
\end{minipage}
\caption{Relative error $\epsilon_{NN}(\boldsymbol{\mu})$ for velocity and pressure fields computed on $\Xi_{te}$, left and right, respectively.}
\label{fig:err_nn_ns_bif_p}
\end{figure}

We chose the hyper-parameters for the POD-NN technique as: $L_K = 2$ layers, $n_b = 18$ and $H_K = 15$, for both velocity and pressure networks.
Considering the testing dataset $\Xi_{te}$ as an equispaced grid of $N_{te} = 301 \times 11$ points in $\Pa$, we are able to reconstruct the wall-hugging behaviour for $11$ different geometries. As regards the accuracy, we show in Figure \ref{fig:err_nn_ns_bif_p} the error $\epsilon_{NN}(\boldsymbol{\mu})$ for the two components of the solution.
Even for this more complex multi-parameter test case we obtain good generalization of the network, since the maximum and the mean relative errors for the velocity field over $\Xi_{te}$, are $\epsilon^{\max}_{NN} = 0.06250$ and $\overline{\epsilon}_{NN} = 0.01181$, respectively. Similar results hold for the pressure field, where we have $\epsilon^{\max}_{NN} = 0.11417$ and $\overline{\epsilon}_{NN} = 0.01150$.

As concerns the RB approximation results, we have $\epsilon^{\max}_{RB} = 0.75538$ and $\overline{\epsilon}_{RB} = 0.01297$ for the velocity field and $\epsilon^{\max}_{RB} = 0.69706$ and $\overline{\epsilon}_{RB} = 0.00984$ for the pressure field.

While the mean errors for the RB and POD-NN techniques are comparable, we observe that the maximum error yields very different result. In fact, still relying on a nonlinear solver during the online phase, the RB technique fails to convergence to the solution (even employing a reduced order continuation method), and provide a bad approximation of the solution especially near the bifurcation point. On the contrary, since the POD-NN strategy is not based on the explicit formulation of the model, it does not encounter any convergence issue.

As regard the computational time, since we are now considering a greater number of basis functions all the efficiency of the reduced model is lost. In fact, each RB or HF solution requires almost $t_{RB} = t_{HF} =10$(s). This means that if we want a complete bifurcation diagram with e.g.\ $n = 301$ values for $\mu$ and $m = 151$ points for $w$, we need  45451 solutions, thus more then 5 days of computational time using either the RB or the HF solver. On the contrary, a single evaluation of the network, which provides the coefficients for the POD-NN approximation, takes only $t_{NN} = 1.2 \cdot 10^{-5}$(s), enabling a speed-up of almost  $8 \cdot 10^{5}$.

Even if we were able to considerably reduce the computational burden to obtain the reduced coefficients, we still have to project them onto the FE space to build the point-wise based bifurcation diagrams in Figure \ref{fig:bifurcation_param}. However, when considering multi-parameter geometric context for bifurcating phenomena, the main goal is to understand the branching behaviour and the location of the critical points, rather then the output.

For these reasons, we applied the technique introduced in Section \ref{sec:redmanbbd} to recover non-intrusively a guess on the location of the bifurcation points for the model.

\begin{figure}
\centering
\includegraphics[height=4.4cm]{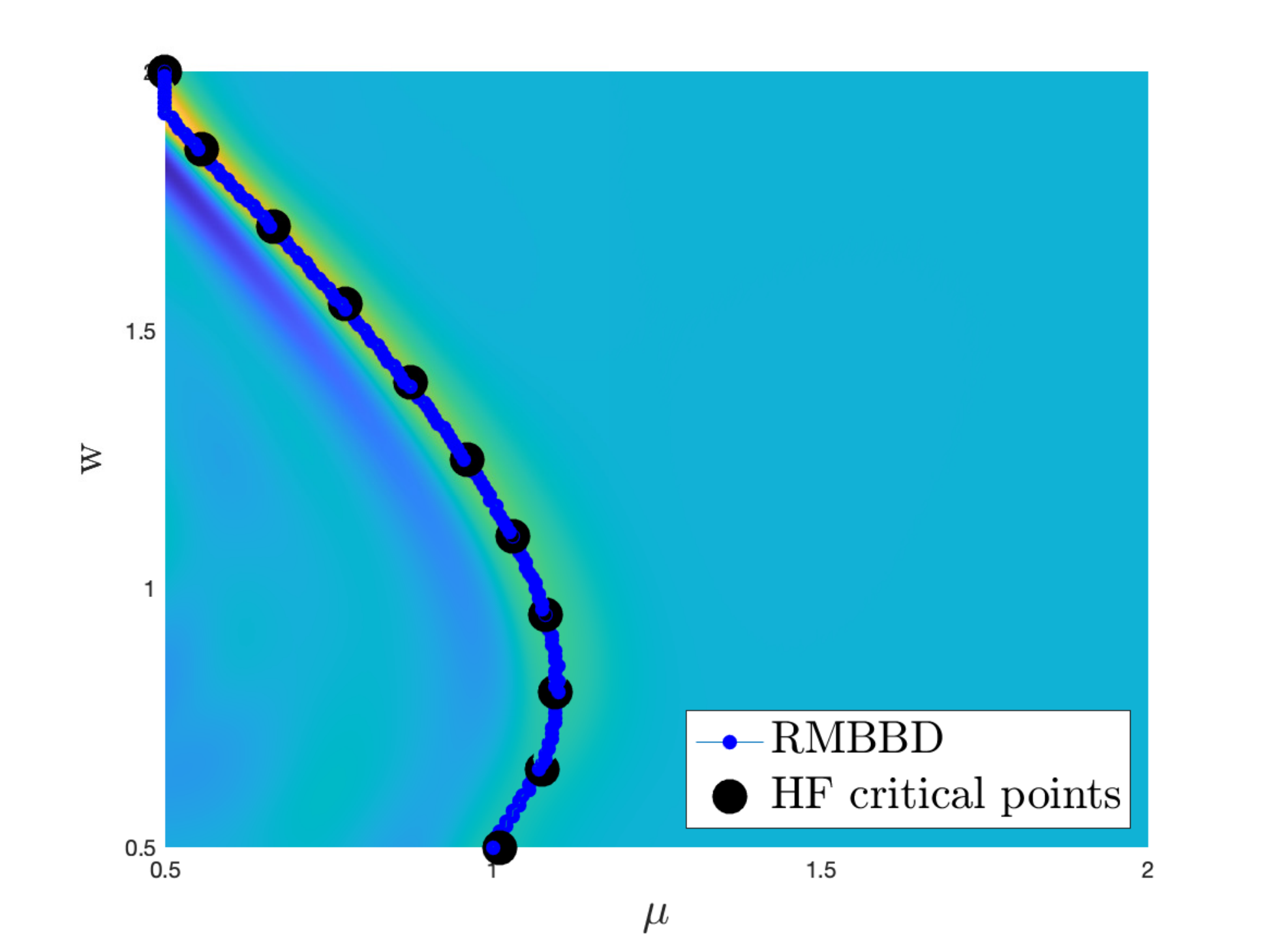}\hfill
\includegraphics[height=5cm]{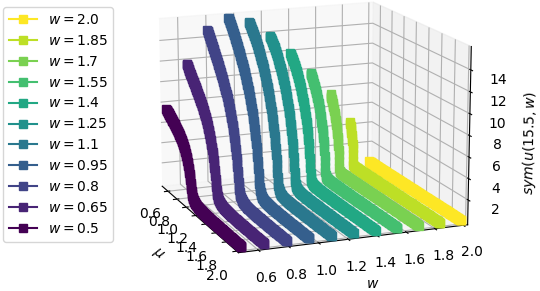}
\caption{\textit{Left}: Reduced manifold based bifurcation diagram for the Coanda model with geometrical parametrization. The black dots represent the high-fidelity critical points, while the blue line corresponds to the maximum curvature obtained trough the POD-NN coefficients. \textit{Right}: Corresponding HF bifurcation diagram obtained through point-wise evaluations of FE solutions.}
\label{fig:curvature}
\end{figure}

Thus, let us consider the extended parameter space $\Pa = [0.5, 2.0]^2$ where the grid $G$ is determined by $n = 301$ and $m = 151$ equispaced points in $\Pa$.
In Figure \ref{fig:curvature}, we plot the output of Algorithm \ref{alg:03}, where the
black dots stand for the exact (high-fidelity) critical points for eleven equispaced geometries, while the blue line is the approximated location for each value $w^{(j)}$ in $G$.
The curvature of the manifold, obtained entirely through the POD-NN coefficients, is able to detect the critical viscosities at which a sudden change in the weighting of the reduced basis functions occurs. As we expected, the evolution of the critical line is not obvious and could represent a fundamental tool to obtain a refined approximation near the bifurcation point for prescribed geometries. Furthermore, this approach could be used to design geometries which can sustain maximum values for viscosity (or for the load in a buckling context \cite{pichipatera}) without undergoing bifurcating phenomena.
We note that in this setting, we are able to recover the critical points behaviour with a relative $l^2$ error of order $10^{-2}$. Moreover, even the non-bifurcating regime near the larger inlet's width values was properly detected by our technique, indicating (by convention) a critical value $\mu = 0.5$.

\subsection{Cavity flow in a triangular domain}
In this section, we investigate the bifurcating phenomena originating in cavity flow problems in a triangular domain. This class of models is of importance in practical applications, as well as benchmark test cases for numerical solver due to their intrinsic complexity. In particular, several studies focused on triangular cavities, investigating: the laminar to chaotic transition \cite{an_bergada_mellibovsky_2019}, the flow behaviour w.r.t.\ grid refinement for different (fixed) geometries \cite{erturk_gokcol_2007} and the 3D context concerning a real life application \cite{gonzalez_ahmed_kuhnen_kuhlmann_theofilis_2011}. For a comprehensive literature, we direct the interested reader to the great historical introduction in \cite{an_bergada_mellibovsky_2019}.
Despite this, a complete knowledge of the dynamics is far from known, as the prediction of recirculating eddies is more involved w.r.t.\ the standard square cavity \cite{erturk_gokcol_2007} and a new steady (bifurcating) state was  recently discovered \cite{an_bergada_mellibovsky_2019}.

Within this complex setting, we aim at investigating the flow's vortex pattern while varying the viscosity and the domain itself.
Thus, we consider a parametrized context involving the advection dominated regime and the geometric parametrization of the triangular cavity in a multi-parameter context. This way, we describe for the first time the vortex pattern behaviour connected to the existence of a critical width angle for the domain.

\begin{figure} [htbp]
\centering
\def\svgwidth{0.4\linewidth}
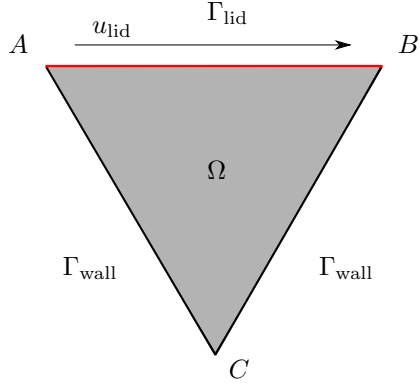
\caption{Domain $\Omega$ representing the triangular cavity.}
\label{fig:tricavity}
\end{figure}

We consider the triangular domain $\Omega$ in Figure \ref{fig:tricavity} identified by the vertices $A = [a_x, a_y]$, $B = [b_x, b_y]$ and $C = [c_x, c_y]$.
Using the same Navier-Stokes model presented in Section \ref{subsec:ns_cha}, we aim at describing the behaviour of the flow inside the cavity when varying its physical and geometric features. As concerns the boundary condition, we chose the following

\begin{equation*}
\begin{cases}
u = u_{\text{lid}}  \quad &\text{on} \ \Gamma_{\text{lid}}, \\
u = 0  \quad &\text{on} \ \Gamma_{\text{wall}}, \\
\end{cases} \quad \text{where} \quad
 u_{\text{lid}} = \begin{bmatrix} 1\\ 0 \end{bmatrix},
\end{equation*}
expressing: no-slip Dirichlet BC on the physical walls $\Gamma_{\text{wall}}$ and a horizontal velocity Dirichlet BC at the lid $\Gamma_{\text{lid}}$.
Moreover, since there are no Neumann conditions, we have to impose, e.g.\ on the bottom vertex $C$, a zero value for the pressure field.

To have a coherent structure, we chose here to remain within the velocity-pressure variables pair as unknowns, instead of modelling the flow by means of the streamfunction-vorticity formulation that usually appears in this context.
%

\subsubsection{Equilateral geometry toy problem}
As a first study case, we investigate the cavity flow in the  one-dimensional parameter space, studying the behaviour of  both the RB and the POD-NN methodologies for the approximation of the solution when varying the viscosity parameter $\mu$.
Before going towards advection dominated regimes, we considered here the parameter space $\Pa = [10^{-3}, 1]$, corresponding to a maximum value of the Reynolds number equal to 1000, for the equilateral geometry in Figure \ref{fig:tricavity}, where $A = [-\sqrt{3}, 1]$, $B = [\sqrt{3}, 1]$ and $C = [0, -2]$.
As concerns the offline phase, we rely again on the Taylor-Hood FE pair, with the fine mesh discretization, resulting in $\N = 41931$ degrees of freedom. Compared to the previous test case we are using a much more refined grid, which is essential to capture the boundary layers, improve the numerical stability controlling the Peclet number and properly approximate step gradients near the corners of the domain.

We chose the hyper-parameters for the RB/POD-NN approximations as follows.
Having a one dimensional parameter space corresponds to an input dimension for the network $P = 1$, while the output layer has a dimension equal to the number of reduced basis considered $N_ u = N_p = 14$, selected by the POD with tolerance $\epsilon_{POD} = 10^{-9}$. We note that for the velocity field we have to add the lifting basis function to allow for the non-homogeneous Dirichlet condition on the moving wall $\Gamma_\text{lid}$.

The network is constructed using $H_K = 15$ hidden neurons for each of the $L_K = 4$ layers, the initial learning parameter is fixed to $\eta_0 = 5 \times 10^{-2}$ with square root learning rate schedule in \eqref{learning_decay}. We train and validate the network with a batch size $n_b = 24$, which corresponds to $N_{train} = 576$ training points.
Such points are selected through a logarithmically equispaced distribution on $\Pa$ and then divided into training and validation sets, following the rule in Section \ref{sec:ann}, for cross-validation purposes. The choice of the sampling's feature is motivated by the fact that slight changes for $\mu$ have a greater importance when the viscosity is small.
For the online simulation $\Xi_{te}$ is chosen as a logarithmically equispaced sample of 250 points in $\Pa$.

\begin{figure}[b]
\begin{minipage}{0.45\textwidth}
\centering
\includegraphics[width=6.5cm]{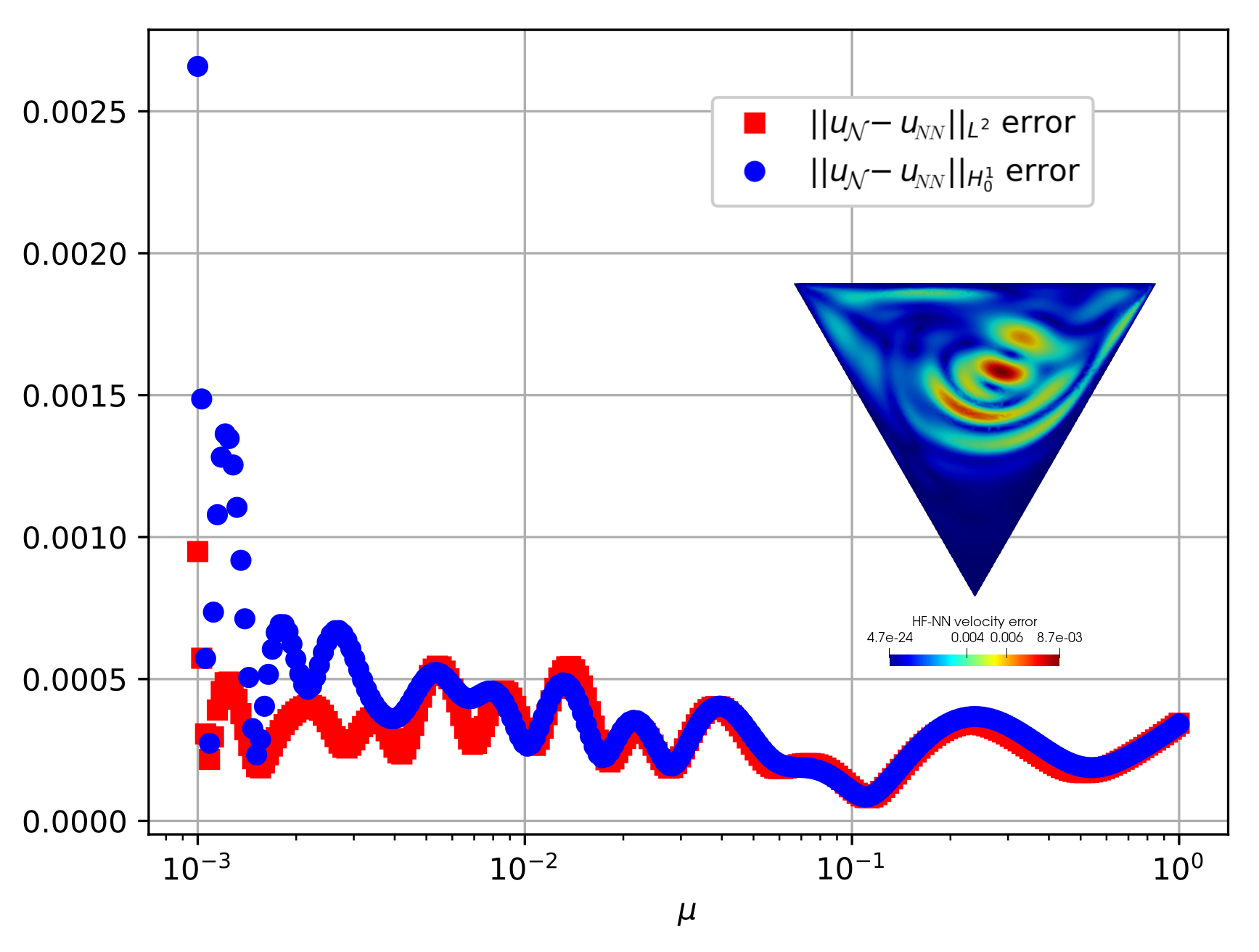}
\end{minipage}\quad\quad
\begin{minipage}{0.45\textwidth}
\centering
\includegraphics[width=6.5cm]{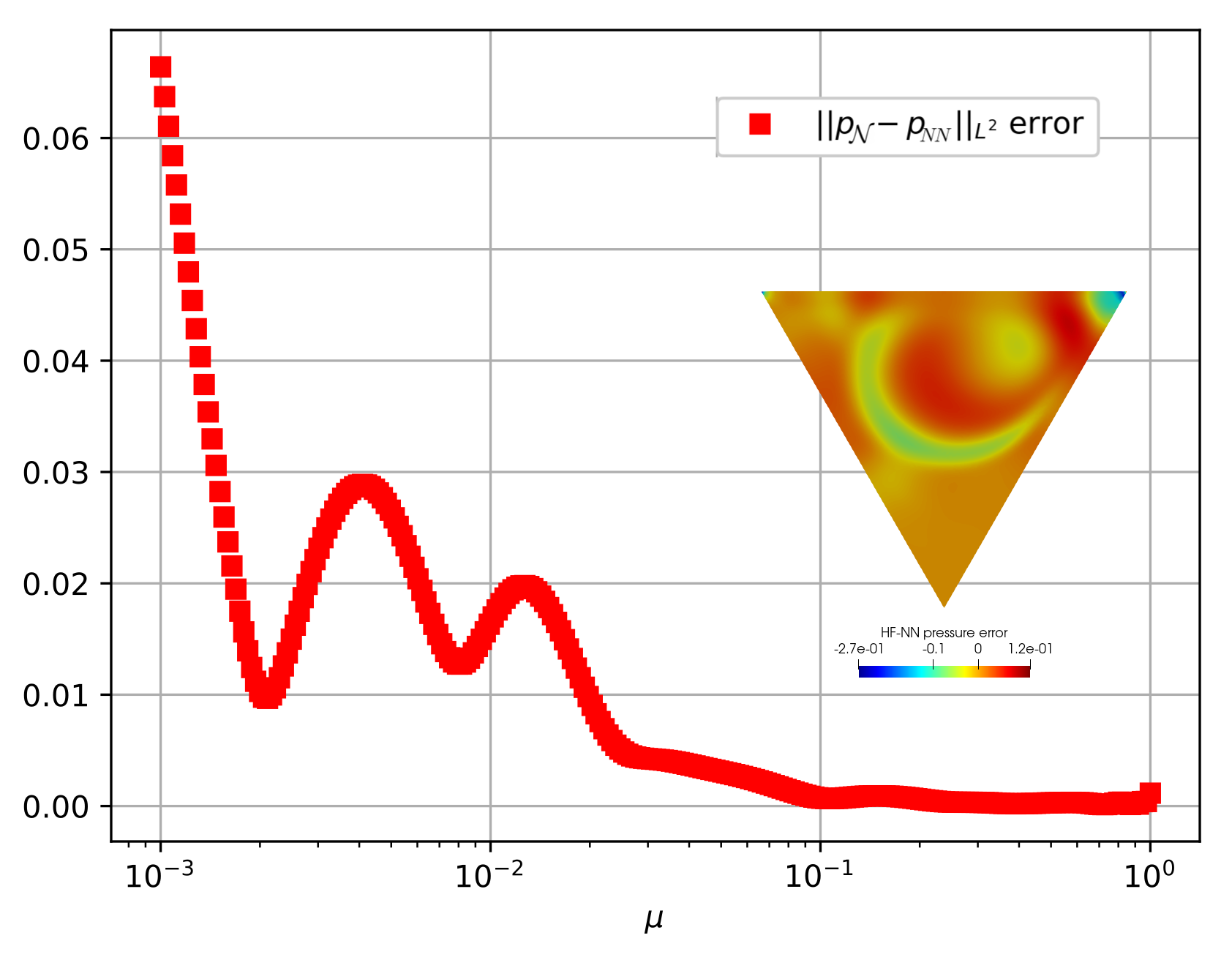}
\end{minipage}
\caption{Relative error $\epsilon_{NN}(\boldsymbol{\mu})$ for velocity and pressure computed on $\Xi_{te}$, left and right respectively, with reference fields at $\mu = 0.001$.}
\label{fig:err_nn_tri}
\end{figure}

\begin{figure}[t]
\begin{minipage}{0.45\textwidth}
\centering
\includegraphics[width=6.5cm]{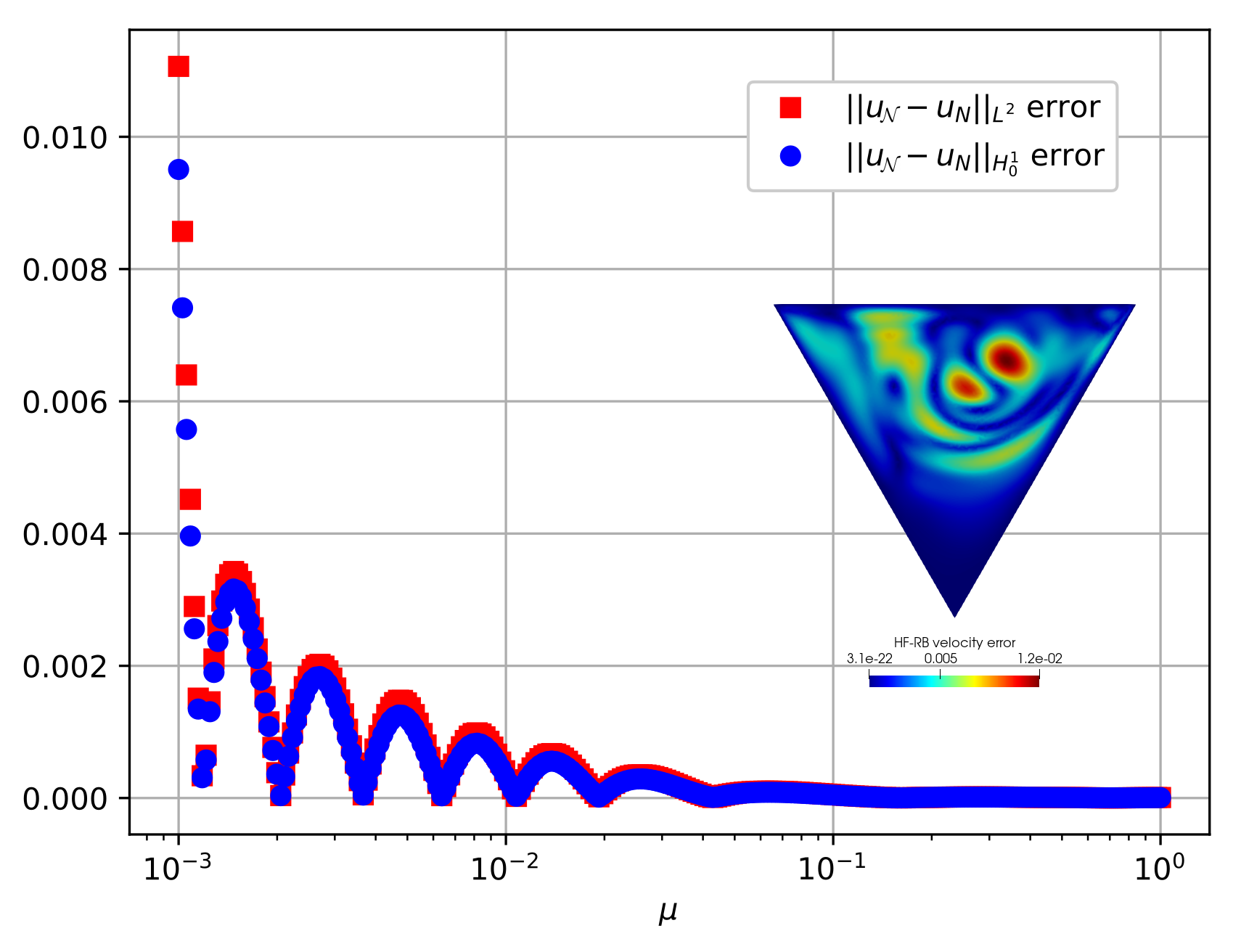}
\end{minipage}\quad\quad
\begin{minipage}{0.45\textwidth}
\centering
\includegraphics[width=6.5cm]{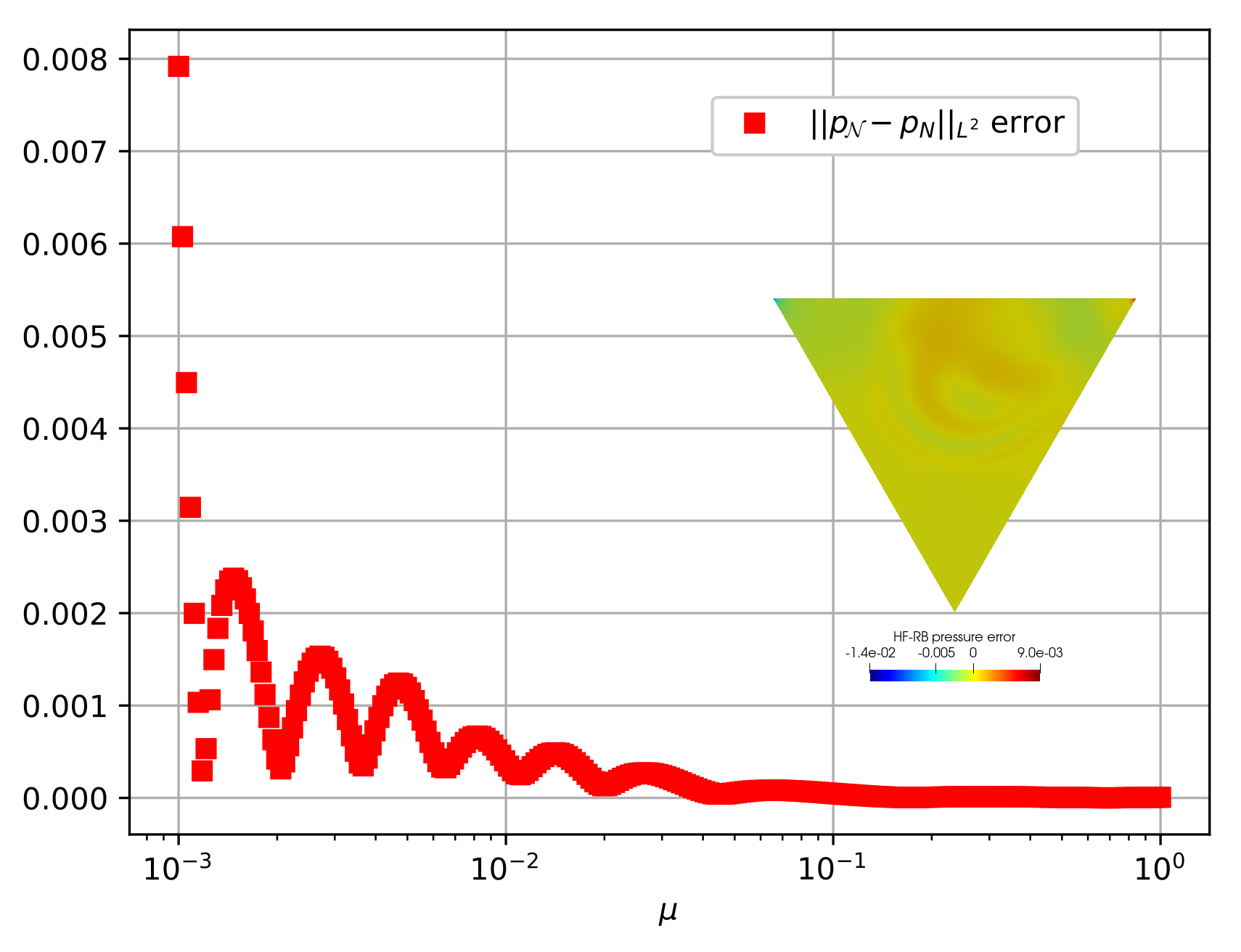}
\end{minipage}
\caption{Relative error $\epsilon_{RB}(\boldsymbol{\mu})$ for velocity and pressure computed on $\Xi_{te}$, left and right respectively, with reference fields at $\mu = 0.001$.}
\label{fig:err_rb_tri}
\end{figure}

As shown in Figure \ref{fig:err_nn_tri}, the maximum and mean relative error over the testing dataset for the velocity field are $\epsilon^{\max}_{NN} = 0.00265$ and $\overline{\epsilon}_{NN} = 0.00037$, respectively. Similar result holds for the pressure field, where we have $\epsilon^{\max}_{NN} = 0.06637$ and $\overline{\epsilon}_{NN} = 0.01091$.
In comparison we show the RB relative error behaviour on $\Xi_{te}$ for the velocity and pressure field in Figure \ref{fig:err_rb_tri}. In both Figures \ref{fig:err_nn_tri} and \ref{fig:err_rb_tri} the error fields are reported for the minimum viscosity value $\mu = 0.001$.
%

We notice that the POD-NN approximation of the velocity field outperforms the one obtained by RB. This is due to the fact that we are employing a low number of basis functions and, again, within the POD-NN approach we are learning from the best possible solution in the RB space, i.e.\ the projection of the snapshots. The same does not hold for the recovery of the pressure field that seems to be more sensitive to the learning procedure.

We note that a much lower number of training points is sufficient to reach a good accuracy of the neural network approximation. As an example, employing only $N_{train} = 36$ snapshots, we are able to obtain a max/mean relative POD-NN error of order $10^{-2}$.

\begin{figure}
\centering
\includegraphics[width=4.2cm]{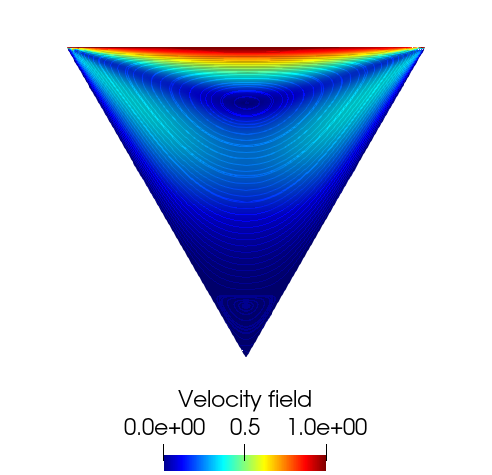}\hspace*{-0.5cm}
\includegraphics[width=4.2cm]{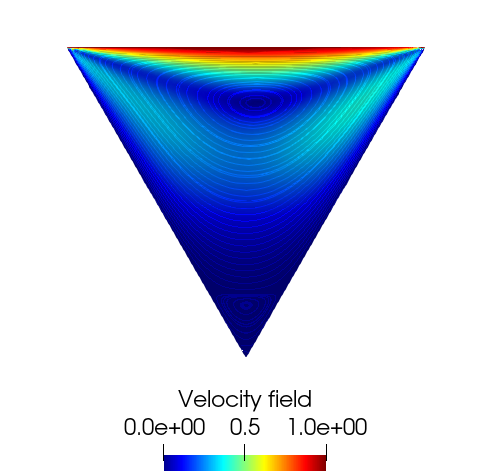}\hspace*{-0.5cm}
\includegraphics[width=4.2cm]{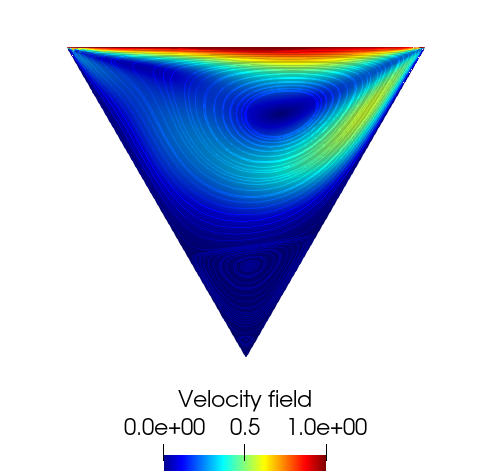}\hspace*{-0.5cm}
\includegraphics[width=4.2cm]{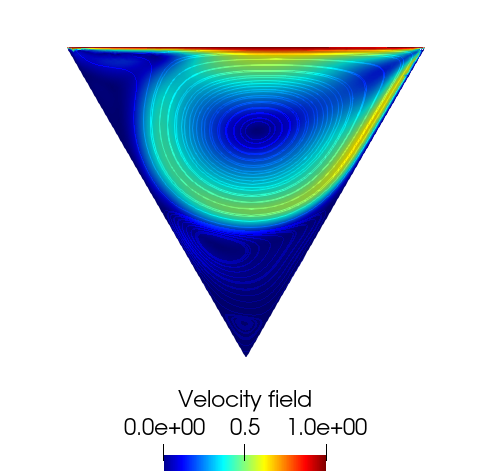}
\caption{High-fidelity velocity fields for the Navier-Stokes system in the triangular cavity for $\mu =  \{1, 0.1, 0.01, 0.001\}$, from left to right.}
\label{fig:TRI_sol_hf_v}
\end{figure}

\begin{figure}
\centering
\includegraphics[width=4.2cm]{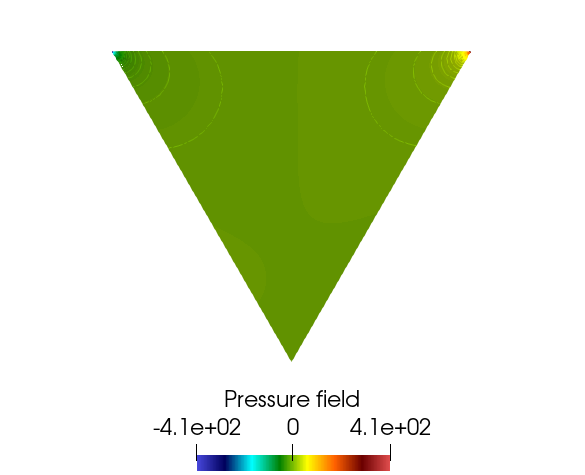}\hspace*{-0.5cm}
\includegraphics[width=4.2cm]{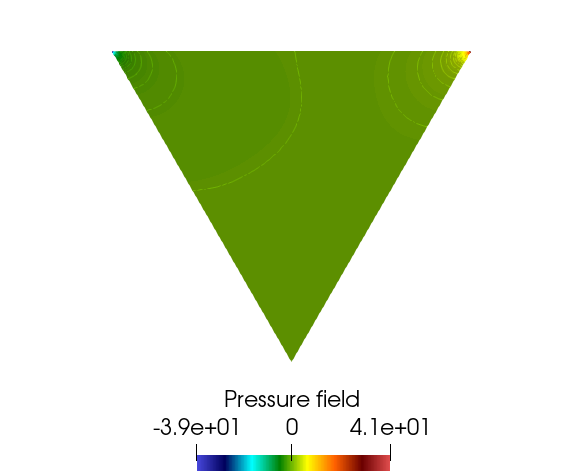}\hspace*{-0.5cm}
\includegraphics[width=4.2cm]{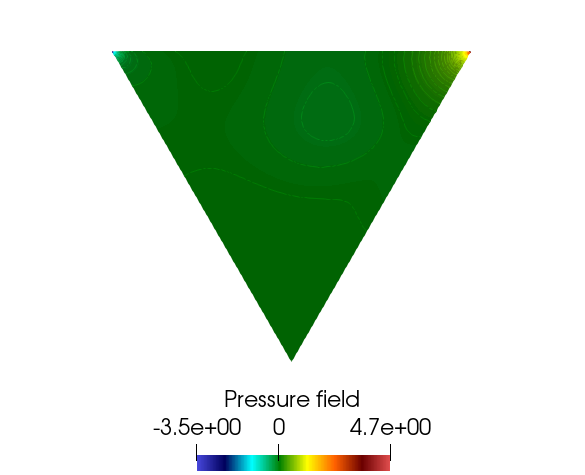}\hspace*{-0.5cm}
\includegraphics[width=4.2cm]{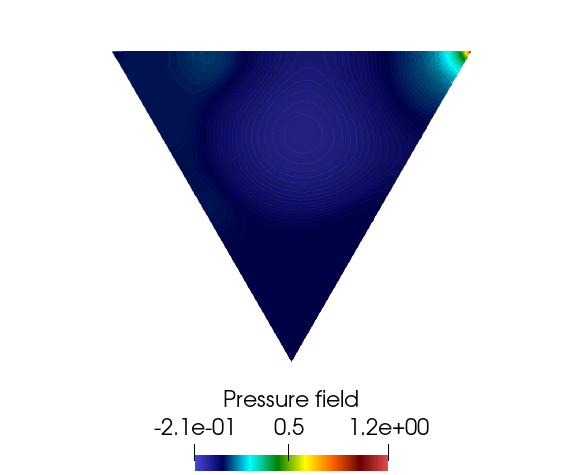}
\caption{High-fidelity pressure fields for the Navier-Stokes system in the triangular cavity for $\mu =  \{1, 0.1, 0.01, 0.001\}$, from left to right.}
\label{fig:TRI_sol_hf_p}
\end{figure}

To understand the flow behaviour w.r.t.\ the parametrization, we show in Figures \ref{fig:TRI_sol_hf_v} and \ref{fig:TRI_sol_hf_p} some representative high-fidelity velocity and pressure fields solutions obtained for different values of the viscosity $\mu = \{1, 0.1, 0.01, 0.001\}$.

Having described the flow behaviour in this relatively simple setting, two questions naturally arise: (i) What happens for different geometries? (ii) How far can we move towards the advection dominated regime? In the next section we will consider a global framework combining these to aspects, while trying to discover the hidden dynamics.

\subsubsection{Bifurcating regimes with parametrized cavities}

Let us begin this section by considering the effect of increasing the Reynolds number on different triangular cavities. To do so, we consider the multi-parameter setting where the viscosity varies in $\Pa = [2\cdot 10^{-4}, 10^{-1}]$, and the vertex $C$ has parametric coordinates $[c_x, c_y] = [\mu_1, \mu_2] \in [-0.5, 0.5] \times [-0.75, -0.25]$. Thus, we redefine the cavity as the triangle identified by the vertices $A = [-0.5, 0], B = [0.5, 0]$ and $C =  [\mu_1, \mu_2]$. Within this low viscosity context we reach the advection dominated regime up to a maximum $Re = 5000$, while the HF degrees of freedom w.r.t.\ the refined mesh increased up to $\N = 283827$.

\begin{figure}
\centering
\includegraphics[width=6cm]{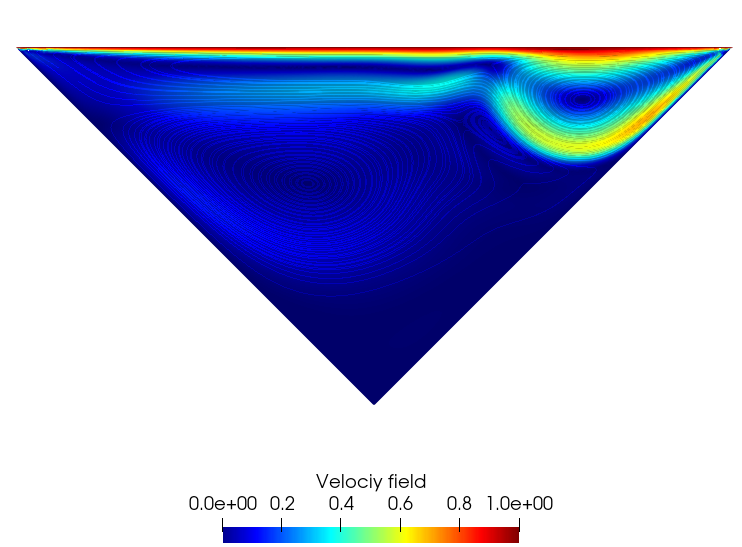}\quad\quad\quad\quad
\includegraphics[width=6cm]{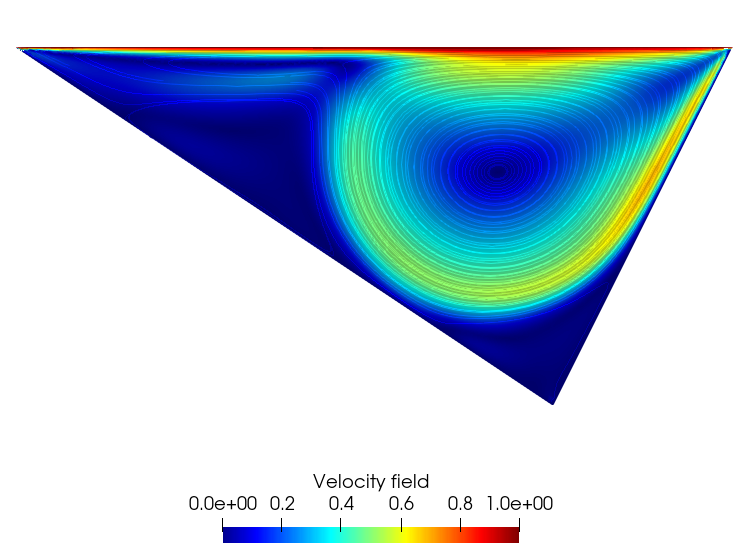}
\caption{POD-NN velocity fields for the Navier-Stokes system in the triangular cavities for $\bmu^1 = (2\cdot 10^{-4}, 0, -0.5)$ and $\bmu^2 = (2\cdot 10^{-4}, 0.25, -0.5)$, left and right, respectively.}
\label{fig:TRI_sol_isononiso}
\end{figure}

In Figure \ref{fig:TRI_sol_isononiso} we show the velocity fields with fixed viscosity for the two cavities characterized by $\bmu^1 = (2\cdot 10^{-4}, 0, -0.5)$ and $\bmu^2 = (2\cdot 10^{-4}, 0.25, -0.5)$. It is clear that even varying slightly a single component of the geometric parametrization, the flow exhibits very different behaviours. In particular, the flow inside the isosceles cavity (related to $\bmu^1$) shows the existence of a vortex attaching to the top-right vertex B, being pushed by the one formed near vertex C. As concerns the configuration corresponding to $\bmu^2$, we observe that the primary vortex is now occupying the whole cavity, attaching to the opposite wall w.r.t.\ the upstream vertex and resulting in an entirely different pattern flow. This suggests the existence of a critical value for the angles' width which determine if have the attaching behaviour, or not.

Before presenting the general framework to study the complex phenomena linked to the geometric parametrization, let us give more insights into the solution behaviour w.r.t.\ the viscosity value $\mu$.

\begin{figure}
\centering
\includegraphics[width=9cm]{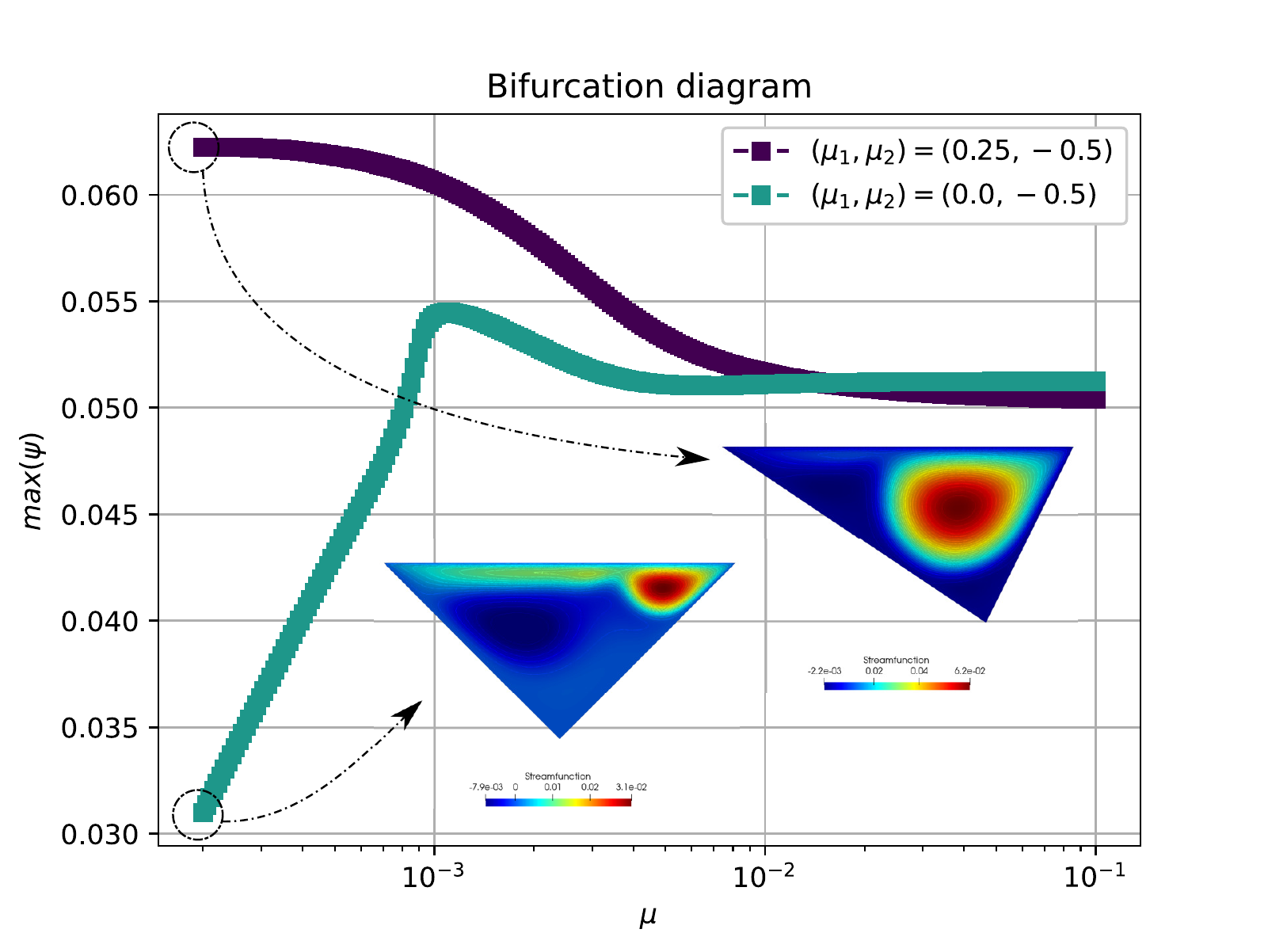}
\caption{Bifurcation diagram w.r.t.\ the maximum value of the streamfunction $\psi$ for the Navier-Stokes system at $\mu = 2 \cdot 10^{-4}$ in the triangular cavities with $(\mu_1, \mu_2) = (0, -0.5)$ and $(\mu_1, \mu_2) = (0.25, -0.5)$.}
\label{fig:bif_full_comp}
\end{figure}

We show in Figure \ref{fig:bif_full_comp} the behaviour of the solution manifold w.r.t.\ the maximum value of the streamfunction $\psi$ over the two cavities, corresponding to $\bmu^1$ and $\bmu^2$. While the output for the non-isosceles configuration has a smooth behaviour, indicating the formation and spreading of the unique central vortex, the situation for the isosceles cavity is entirely different. In fact, we observe a sudden change in the output, linked to the movement of the vortex towards the top-right vertex.
This behaviour suggests that: (i) the flow profile is highly dependent on the domain, and (ii) it seems that, for the isosceles cavity, we are looking at a bifurcating phenomenon while varying $\mu$ near $10^{-3}$.
In fact, the latter, could be connected to the other stable steady solution found in \cite{an_bergada_mellibovsky_2019} departing from $Re \approx 4908$.

\begin{figure}
\centering
\includegraphics[width=0.48\textwidth]{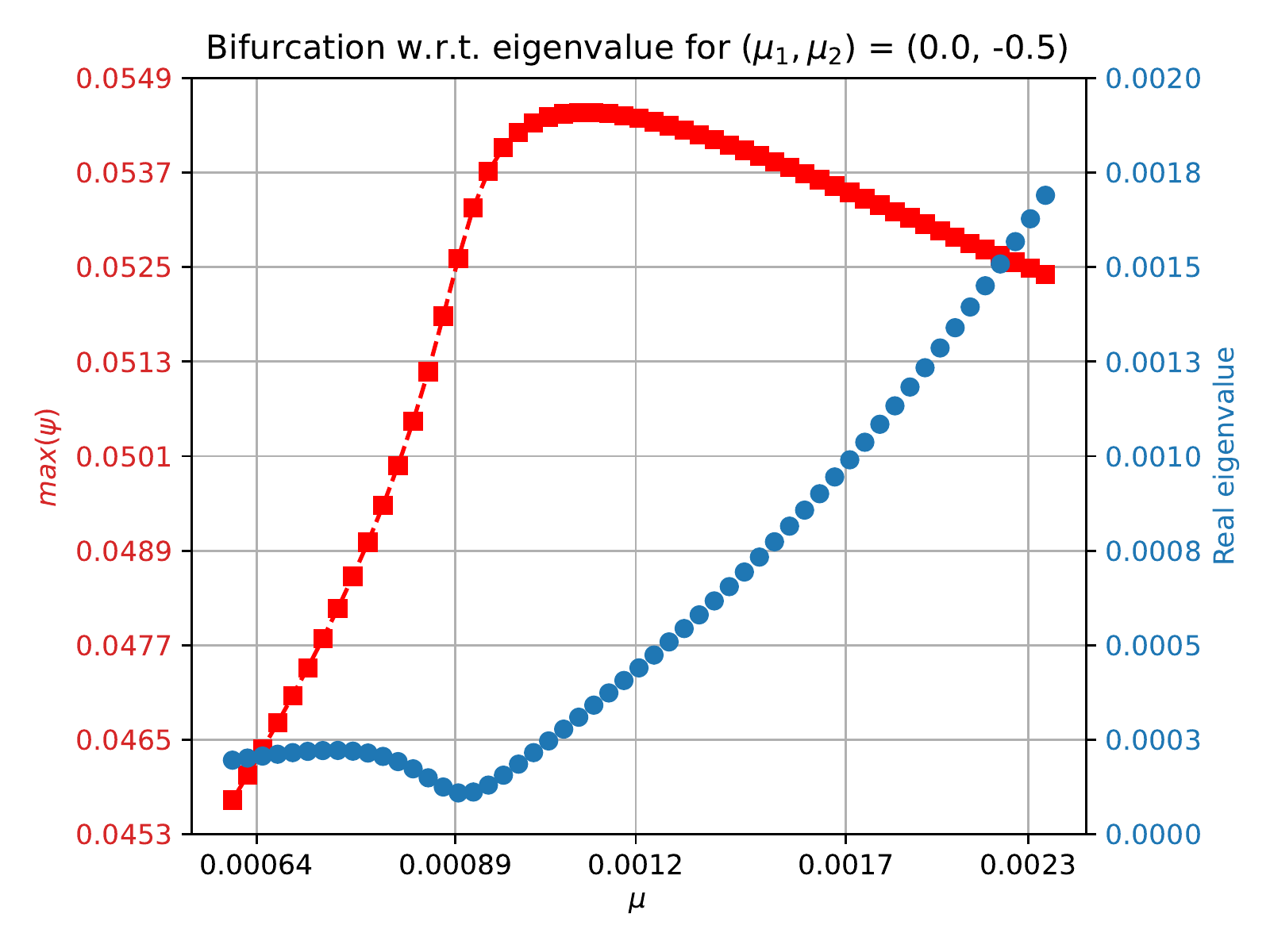}\quad
\includegraphics[width=0.48\textwidth]{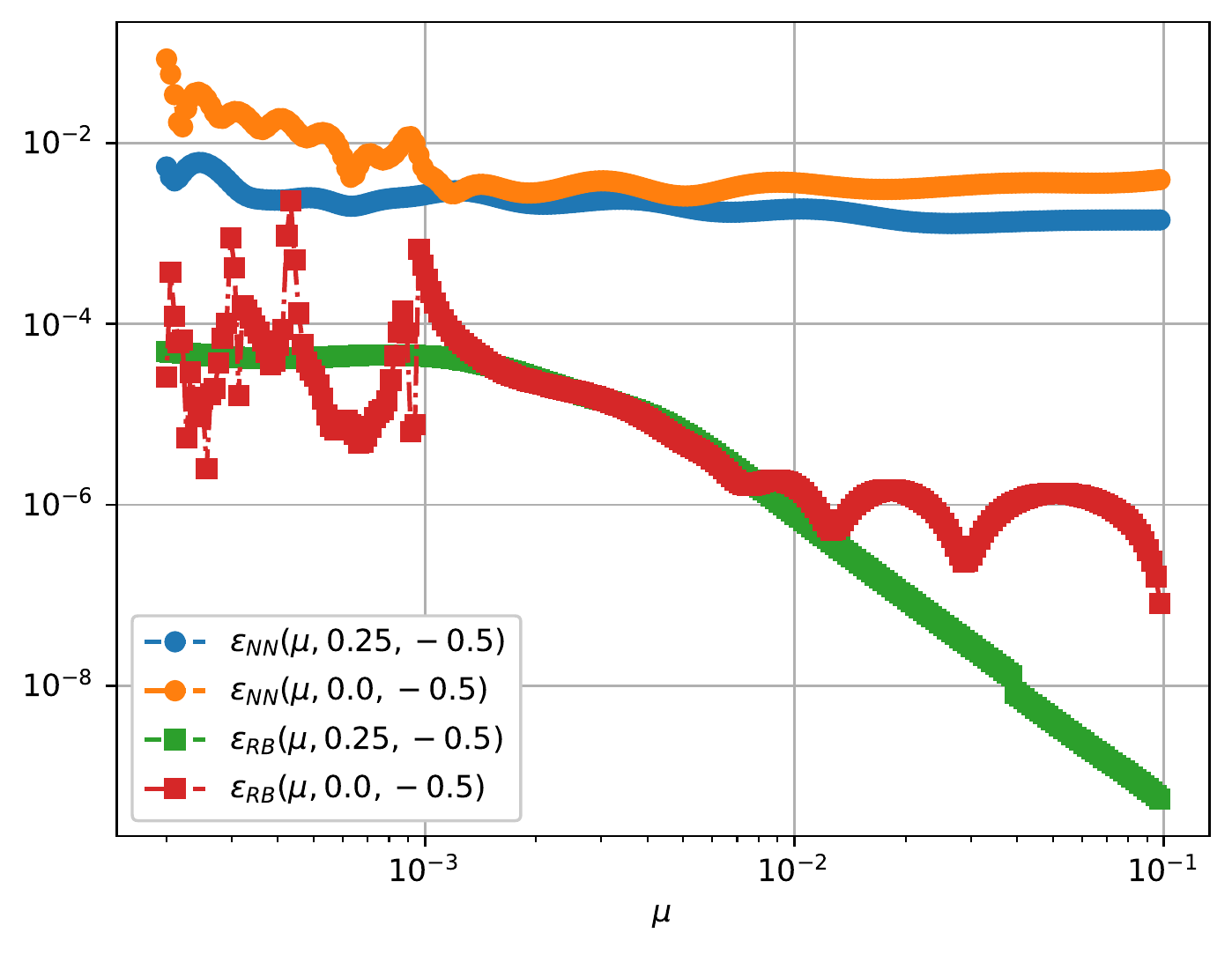}
\caption{\textit{Left}: Comparison between bifurcating branch and eigenvalue dynamics for the isosceles cavity. \textit{Right}: RB and POD-NN log-relative errors for the two configurations w.r.t.\ $\mu$ for the velocity field.}
\label{fig:eigenandrbnn}
\end{figure}

To further investigate the isosceles flow dynamics, we present in Figure \ref{fig:eigenandrbnn} a plot comparing the bifurcation diagram and the evolution of the smallest real eigenvalue of the generalized Navier-Stokes eigenvalue problem. Following the analysis in \cite{pichi_bif, pichistrazzullo} and knowing the connection between stability properties and the sign of the eigenvalues, we observe that in the same interval where the output shows the sudden change, the eigenvalue moves towards zero and bounces back, thus ensuring the flow's stability. Indeed, a change in the sign of the eigenvalue would have denoted a change in the stability features of the solution and so the presence of a new branch. In this case, we have followed the stable one, corresponding to the positive eigenvalue.

To conclude this analysis, we briefly present the RB and POD-NN results concerning the high Reynolds regime for the two cavities. To have a uniform setting we sample $N_{train} = 100$ snapshots, extracting $N_u=N_p=30$ basis functions, and test the intrusive and non-intrusive techniques on $\Xi_{te}$, constructed by 250 logarithmically equispaced points on $\Pa$. As concerns the NN structure, given the increased parametric range and number of basis, we employed $H_K = 60$ neurons for each one of the $L_K = 4$ layers. We can observe in Figure \ref{fig:eigenandrbnn} that, once the RB space is built with a proper number of modes, we obtain good approximation properties for the standard POD-Galerkin approach, for both configurations. Despite this, trying to encode the moving vortex, the RB space for the isosceles domain, when compared with the other one, produces a similar mean error $\bar{\epsilon}_{RB}$ of order $10^{-5}$, but also a much worse maximum error, e.g., we obtain $\epsilon_{RB}^{\max}(\cdot, 0, -0.5) = 0.00227$ and $\epsilon_{RB}^{\max}(\cdot, 0.25, -0.5) = 0.00005$.
The same effect, but mitigated to only one order of magnitude, can be seen for the POD-NN approximation. This is due to the fact that without intrusiveness the equation are not taken into consideration, and thus the mean $\bar{\epsilon}_{NN}$ and max $\epsilon^{\max}_{NN} $ errors share the same order of accuracy. Of course, the better approximation performance of the intrusive RB technique are balanced by the great speedup (of order $10^6$) obtained  through the non-intrusive POD-NN approach.

Having clarified the main differences when looking at two different configurations, now we aim at investigating for which Reynolds numbers this top-right attaching vortex behaviour exists, and its connection to the angles defining the triangular domain.

To obtain a global investigation of the phenomena we consider here the physical/geometric multi-parametrized setting, with $P = 3$, given by $\bmu = (\mu, \mu_1, \mu_2) \in \Pa = [4\cdot 10^{-4}, 10^{-1}] \times [-0.5, 0.5] \times [-0.75, -0.25]$, where $(\mu_1, \mu_2)$ are the coordinates of the vertex C with reference configuration  $\overline{C} = (0, -1)$.
Since we are changing the characteristic of the triangular domain, substantially we expect that this will influence the flow behaviour. Thus, we halved the minimum viscosity value, considering a maximum $Re = 2500$. While remaining in the advection dominated regime, this avoids divergence of the Newton solver during the offline training. On the one hand, within the geometrically parametrized context we do not want/need to recompute the mesh for each cavity, but on the other hand, to obtain good approximations at the HF level, we have to consider a properly refined grid, which has to be regular especially near the corners.

Moreover, in this case we have a unique subdomain, corresponding to the entire cavity, where we need to define an affine transformation map. Following the analysis in Section \ref{sec:ann_multiparam}, we define the transformation matrix as $\mathbb{B}(\bmu) = \left[1, \, -\mu_1; 0, \, -\mu_2\right]$ and express the weak formulation of the problem in the reference configuration through the variational forms in \eqref{eq:forms_param}.

\begin{figure}
\centering
\includegraphics[width=9cm]{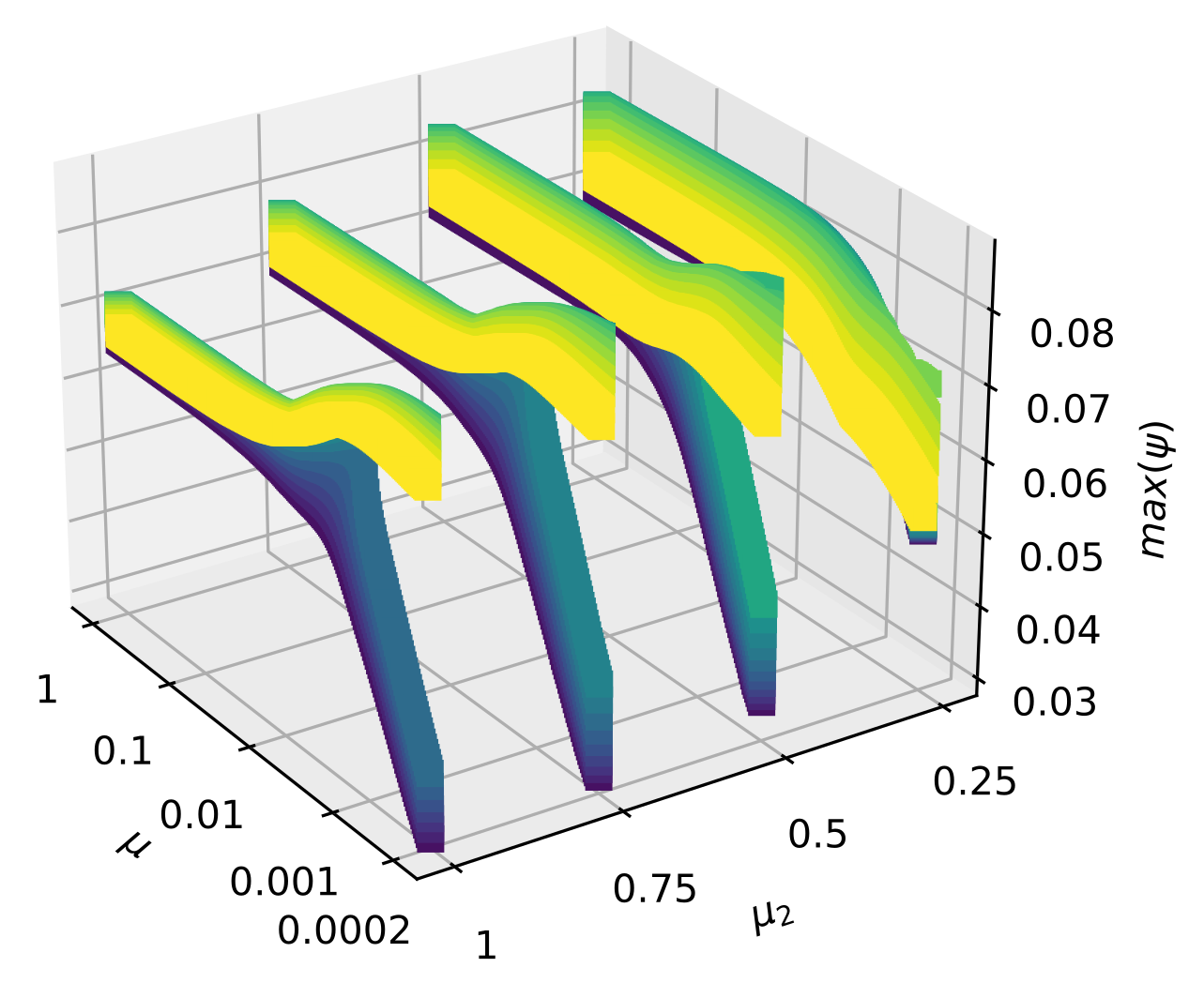}
\caption{4-D bifurcation diagram with the maximum value of the streamfunction $\psi$ as output for the Navier-Stokes system, w.r.t.\ the viscosity $\mu$, the y-coordinate $\mu_2$ of C,  and from violet to green increasing values for the x-coordinate $\mu_1$ of C.}
\label{fig:bif_4d_tri}
\end{figure}

To give an idea of the complexity of the combined effect of high Reynolds number and existence of the critical angles' width, we show in Figure \ref{fig:bif_4d_tri} the ``4-D" bifurcation diagram where we plot the maximum value of the streamfunction $\psi$ w.r.t.\ the viscosity $\mu$ and the y-coordinate $\mu_2$ of C , while with the colors from violet to green denote increasing values for the x-coordinate $\mu_1$ of C.
In Figure \ref{fig:bif_4d_tri} we clearly observe the advection dominated effect, due to the low viscosity regime, depending on the abscissa of vertex C, and the evolution of this phenomenon while varying the height of the triangular cavity. In particular, we have chosen a grid for $(\mu, \mu_1, \mu_2)$ of $250 \times 21 \times 4$ equispaced points, on which we computed the HF solutions. As we have done in Section \ref{sec:ann_multiparam}, we aim at having an even refined  investigation while obtaining substantial a speedup.

Moreover, the different behaviour of the manifold for $\mu_2 = 0.25$. Even if it is not clearly visible from the bifurcation diagram (with this output function), also for this configuration we have both attaching and spreading vortex phenomena while varying $\mu_1$. Thus, we claim that the different dynamics w.r.t.\ the streamfunction is due to the fact that almost all configurations with $\mu_2 = 0.25$ are actually obtuse triangles, minimizing the differences between the two pattern flows.

We are finally ready to study the RB and POD-NN approximations of the high-fidelity results presented up to now.
As concerns the offline procedure, here we sample $\bmu$ in an equispaced grid of $100 \times 11 \times 5$ points in the parameter space $\Pa$.

Instead of fixing the POD tolerance we decide a priori a much higher dimension for the reduced basis space with $N_u = N_p = 150$.
Consequently, we modified the structure of the network for the POD-NN technique as: $L_K = 5$ layers and $H_K = 80$ neurons, for both velocity and pressure networks.
Regarding the online phase we considered the testing dataset $\Xi_{te}$ as 4 random couples $(\mu_1, \mu_2)$ of which we study the evolution for 250 viscosity values $\mu$.

\begin{figure}
\begin{minipage}{0.45\textwidth}
\centering
\includegraphics[width=6.5cm]{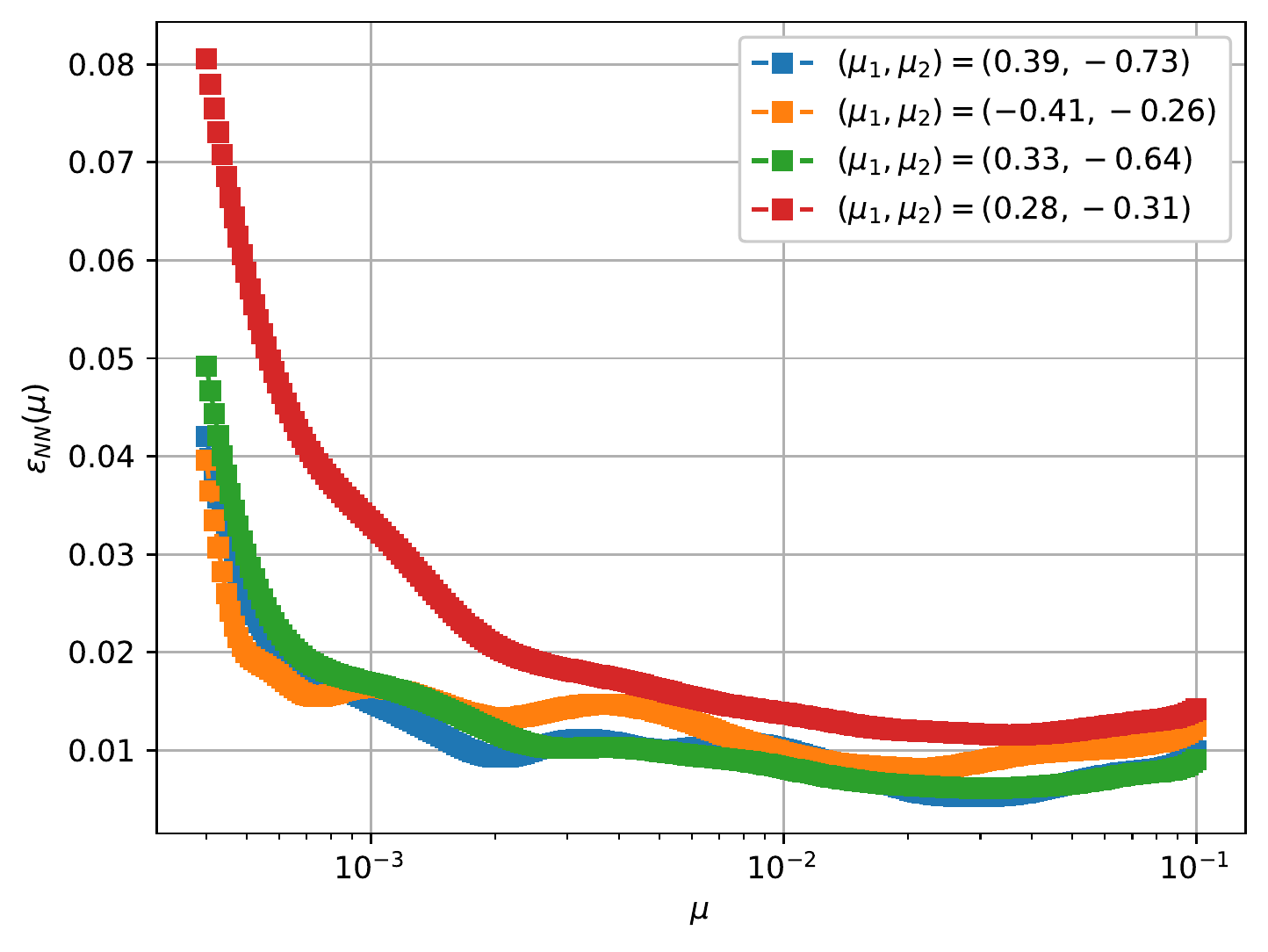}
\end{minipage}
\begin{minipage}{0.45\textwidth}
\centering
\includegraphics[width=6.5cm]{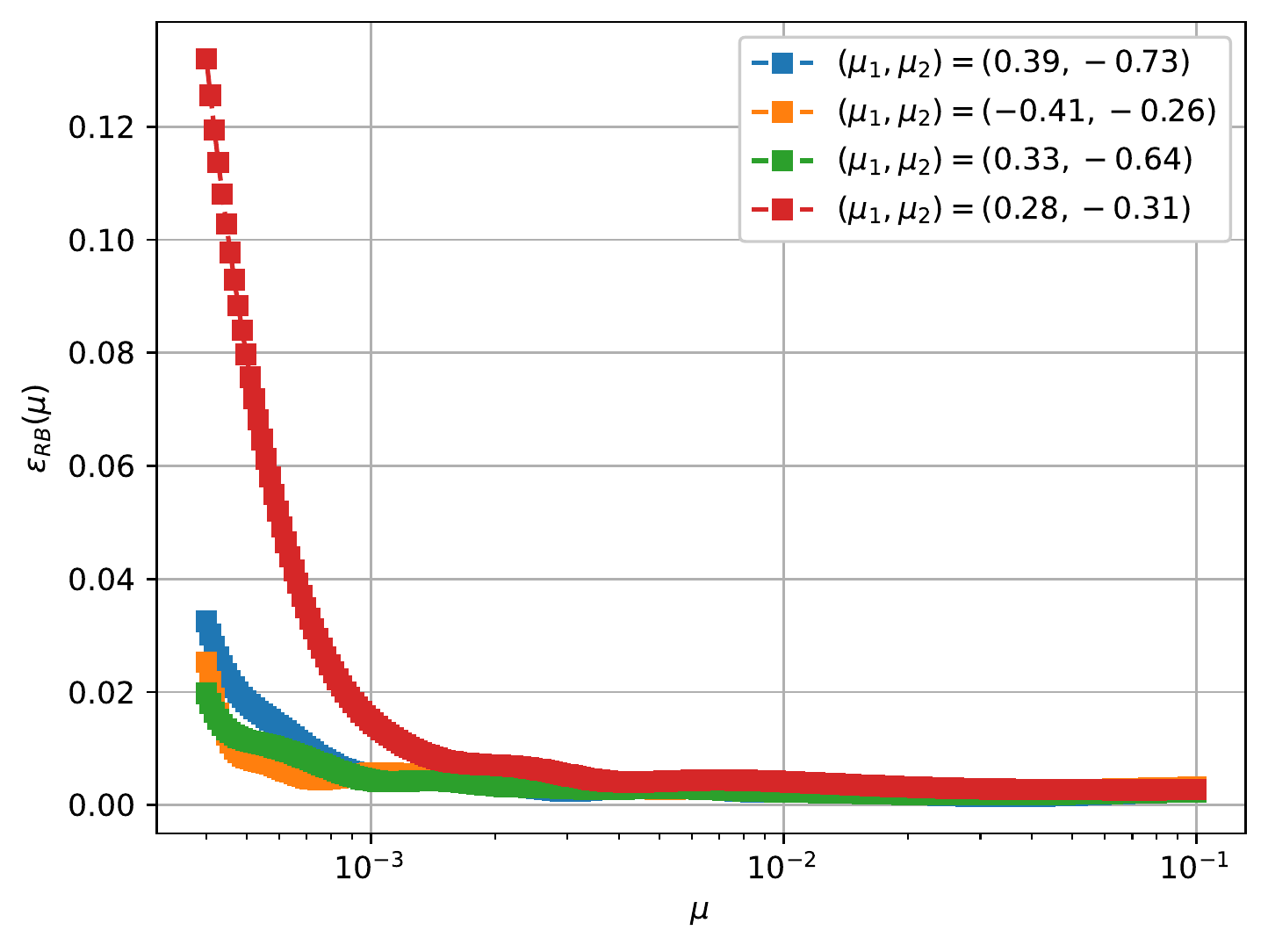}
\end{minipage}
\caption{Relative errors $\epsilon_{NN}(\mu)$ and $\epsilon_{RB}(\mu)$ for the velocity field computed on $\Xi_{te}$, left and right, respectively.}
\label{fig:error_param}
\end{figure}

As concerns the accuracy, we show in Figure \ref{fig:error_param} the velocity relative errors $\epsilon_{NN}(\mu)$ and $\epsilon_{RB}(\mu)$ for the four random geometries.
We were able to obtain good POD-NN approximations, with mean and maximum errors of order $10^{-2}$. Similar accuracy have been obtained by means of standard POD-Galerkin approach, where, as before, if compared with the POD-NN technique we have a better mean error but a worst maximum one. Once again, the main result here is the computational speedup which allows us to evaluate efficiently the reduced coefficients in a  non-intrusive manner. To be more precise, to obtain the evolution in $\mu$ for a single test cavity through the HF technique we need $t_{HF} = 4500(s)$, while the same computation within the POD-NN approach required only $t_{NN} = 0.05(s)$ for the network evaluation. On the contrary, when employing such a great number of basis functions for the RB space, the POD-Galerkin strategy (without hyper-reduction) is no longer advantageous when compared to the Galerkin-FE method.

As already observed in Section \ref{sec:ann_multiparam}, even if we have obtained the reduced coefficients, we still have to rely on the HF degrees of freedom  $\N$ to recover the solutions and extract the bifurcation diagram.
We applied again the strategy developed in Section \ref{sec:redmanbbd} to discover the pattern flow evolution and, in particular, which of the two phenomena, attaching or spreading vortex, we obtain for all the possible parametrized cavities.

\begin{figure}
\begin{minipage}{0.48\textwidth}
\centering
\includegraphics[width=7cm]{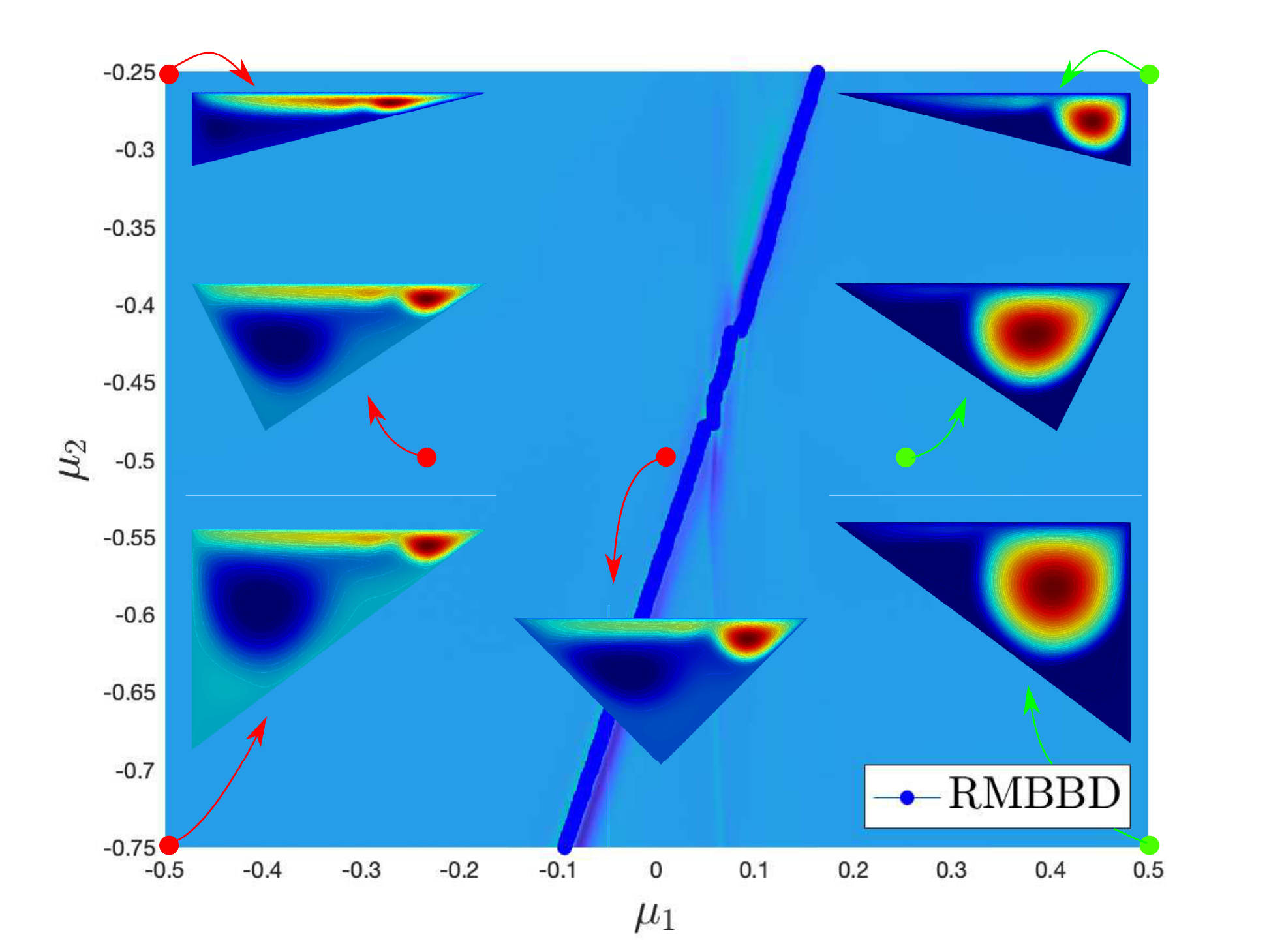}
\end{minipage}
\begin{minipage}{0.48\textwidth}
\centering
\includegraphics[width=7cm]{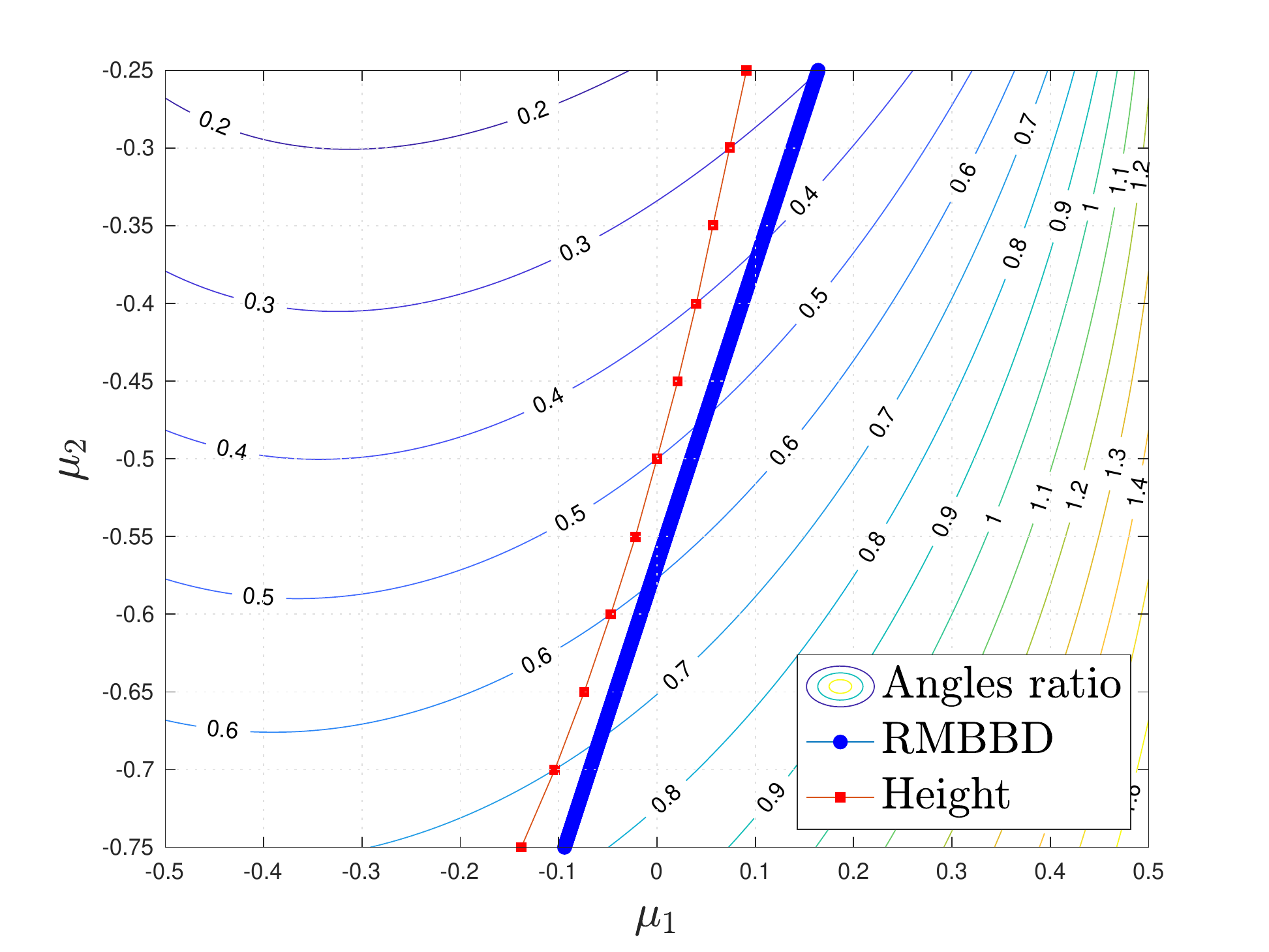}
\end{minipage}
\caption{\textit{Left}: Reduced manifold based bifurcation diagram for the triangular cavity problem with geometric parametrization. The blue line corresponds to the maximum curvature obtained trough the POD-NN coefficients. \textit{Right:} Contour lines of angles ratio comparing the evolution of the RMBBD critical points and height values.}
\label{fig:curvature_tri}
\end{figure}

Let us focus on the high Reynolds regime, corresponding to the lowest viscosity value in the parameter space $\mu = 4 \cdot 10^{-4}$. Given the high sensitivity of the solution  w.r.t.\ the geometric parameters, we chose the grid $G$ as $n = 501$ and $m = 501$ equispaced points in $[-0.5, 0.5] \times [-0.75, -0.25]$.
In Figure \ref{fig:curvature_tri}, we plot the output of Algorithm \ref{alg:03}, where the blue line is representing the approximated locations of the critical parameters, and so the critical angles' width, that serves as separation for the region where we observe attaching vortices (to the left of the line) and the region where we observe spreading ones (to the right of the line).
The curvature of the manifold is able to detect the critical parameters at which a sudden change of the flow solution occurs. We highlight that such analysis, involving $n\times m$ solutions, would not have been possible through standard FE or RB techniques.


Even if the critical line seems to be obvious and due to some rescaling argument concerning the geometrical parametrization, this is not the case. In fact, the computed transformation map cannot be expressed in terms of a single parameter. Moreover, the direction of the inlet flow $u_{lid}$ breaks the symmetry of the cavity problem and make the critical locations less than obvious.
Finally, while trying to explain this phenomenon, we observe that a possible guess for the location of the critical points is given when the ratio between the angles' width of vertices B and C is equal to the height of the cavity, namely $\mu_2$ with changed sign.
To show this, we plot in Figure \ref{fig:curvature_tri} the contour lines of the angles ratio, i.e.\ $\widehat{B}/\widehat{C}$ for every pair $(\mu_1, \mu_2) \in G$. As we see, the blue line given by the RMBBD is very close to the red one, indicating when the angles ratio function assumes the value of the cavity's height $\mu_2$.

\section{Conclusions and perspectives}\label{sec:conc}
In this work, we have discussed the use of an artificial neural network to efficiently recover the evolution of bifurcating phenomena in computational fluid-dynamics problems. We investigated the multi-parametrized setting with varying geometry, analysing its effect on the position of the bifurcation points. In particular, we discussed the physics behind the Coanda effect in a sudden expansion channel and the lid driven triangular cavity flow, providing new insights on recently discovered phenomena. Beyond the discussion of the models, we studied their numerical approximation in the reduced order modelling context, building the basis with a branch-wise POD technique. Moreover, we combined the latter with an artificial neural network, through the POD-NN approach, which provides good approximation properties even when considering such complex behaviours. Finally, we developed a reduced manifold based bifurcation diagram, able to detect in a non-intrusive manner the position of the critical points.

Many possible directions can be further investigated, improving the approximation strategies or focusing on the CFD models.
As concerns the former, one could choose a more advanced NN technique based on autoencoder or physics informed neural networks. Instead, moving the focus onto the bifurcating models, it could be interesting to consider an improved description of the Coanda effect, for example based on a fluid-structure interaction problem, or deepen the fluid-dynamic analysis of the triangular cavity flow, which could undergo more complex phenomena.


\section*{Acknowledgements}
We acknowledge the support by European Union Funding for Research and Innovation -- Horizon 2020 Program -- in the framework of European Research Council Executive Agency: Consolidator Grant H2020 ERC CoG 2015 AROMA-CFD project 681447 ``Advanced Reduced Order Methods with Applications in Computational Fluid Dynamics''. We also acknowledge the PRIN 2017 ``Numerical Analysis for Full and Reduced Order Methods for the efficient and accurate solution of complex systems governed by Partial Differential Equations" (NA-FROM-PDEs), the INDAM-GNCS project ``Tecniche Numeriche Avanzate per Applicazioni Industriali" and the ``GO for IT" program within a CRUI fund for the project ``Reduced order method for nonlinear PDEs enhanced by machine learning".

\bibliographystyle{abbrv}
\bibliography{bib}

\end{document}